\documentclass[10pt]{iopart}
\usepackage{bm,amssymb,graphicx,mathrsfs,color}

\definecolor{dgreen}{rgb}{0,0.7,0}

\usepackage{ulem}

\graphicspath{{FIGURES/}}

\newcommand{\bx}{\mathbf{x}} 
\newcommand{\by}{\mathbf{y}}
\newcommand{\jj}{\mathcal{J}}

\newcommand{\Z}{Z}
\newcommand{\nz}{{L_Z}}
\newcommand{\cE}{\mathcal{E}}
\newcommand{\cF}{\mathcal{F}}
\newcommand{\pin}{p_\mathrm{in}}

\newcommand{\cN}{\mathcal{N}}
\newcommand{\cL}{\mathcal{L}}

\newcommand{\dd}{\mathrm{d}}
\newcommand{\ee}{\mathrm{e}}
\newcommand{\ci}{\mathrm{i}}

\begin{document}

\title{
Zone clearance in an infinite TASEP with a step initial condition
}

\author{
 Julien Cividini$^1$, 
 C\'ecile Appert-Rolland$^1$}

\address{$^1$ Laboratoire de Physique Th\'eorique, b\^atiment 210 Universit\'e Paris-Sud and CNRS (UMR 8627), 91405 Orsay Cedex, France}

\date{\today}

\begin{abstract}
The TASEP is a paradigmatic model of out-of-equilibrium statistical physics, for which many quantities have been computed, either exactly or by approximate methods. In this work we study two new kinds of observables that 
have some relevance
in biological or traffic models. They represent the probability for a given clearance zone of the lattice to be empty (for the first time) at a given time, starting from a step density profile. Exact expressions are obtained for single-time quantities, while 
more involved 
history-dependent observables are studied by Monte Carlo simulation, and partially predicted by a phenomenological approach.
\end{abstract}

\maketitle

Although it has been proposed first for biological
applications~\cite{macdonald_g_p1968}, the Totally Asymmetric Simple Exclusion
Process, or TASEP, has quickly become a paradigmatic model of
out-of-equilibrium statistical physics and of stochastic processes~\cite{derrida1998c,kriecherbauer_k2010}.
It has been
shown to be a minimal model for various transport processes, for example in
biological
systems~\cite{chou_m_z2011,appert-rolland_e_s2015}
or in car traffic~\cite{chowdhury_s_s2000, schadschneider2008b}.
The TASEP consists of point-like particles hopping on a one-dimensional discrete
lattice. Particles hop only to the right, and only if the target site is
empty. Though the model definition is very simple, the model has nevertheless
a very rich
behaviour, exhibiting boundary driven phase transitions.
The TASEP has
mainly been studied with periodic or open boundary conditions, or on an
infinite line.

The role of defects has been extensively studied in the
TASEP~\cite{janowsky_l1992, janowsky_l1994, mallick1996, imamura_s2007,  turci2013, ito_n2014, 
		wang2014, sinha_c2015, sahoo_d_k2015, wang_j_w2017}.
In several applications the defects are
dynamic and may
depend on the local configuration of particles.
A particular case of such defects is given by on- and off-gates.
In the case of pedestrians, such a gate could for example result
from a counterflow or a competing flow temporarily blocking the
access to a bottleneck~\cite{jelic_a_s2012, yuan2008b, appert-rolland_c_h2011c}.
Some processes regulating protein synthesis along
mRNA strands could also be modelled by such on- and off-gates~\cite{turci2013}.
In these two examples, the temporary closure of the gate can
occur only if a certain region (which we shall refer to as the clearance zone)
is void of particles. Once closed, the gate
prevents particles to pass until it opens again. If enough particles accumulate
behind the gate during the closure,
an efficient transient flow takes place
at the opening, with an initial condition that can be approximated by a
step~\cite{jelic_a_s2012}.
Predicting the probability that during this transient,
the clearance zone could be empty - and thus allow for a new closure of the gate - is an open question.
The work presented in this paper can be seen as a
first stage to study this coupled dynamics between closure of the lattice and
positions of the particles.

In this paper, we shall consider the TASEP on an infinite line (or \textit{infinite TASEP}),
with a step initial condition (namely all the sites to the left of a given site are occupied and all the others are empty at initial time).
For the TASEP on an infinite line,
the propagator of the system can be derived exactly by Bethe Ansatz and takes the form of a determinant~\cite{schutz1997}.
Besides, in the case of the initial step condition,
exact results can be obtained, in particular for the statistics of the particle
current, and much effort has been made to obtain the asymptotic behavior of
\textit{e.g.} density profiles or the large deviation function of the current in the limit of large times~\cite{chou_m_z2011, tracy_w2009, baik_l2016}.

The observables we shall be interested in are the
probabilities that a certain zone of the lattice located to the right of the
initial step is empty, or empty for the first time.
These probabilities will be computed either regardless of
the past history or with the requirement that the zone has always been occupied
before.

In section \ref{section:introstep} we define the infinite TASEP
with a step initial condition and outline some useful results from the literature. Section \ref{section:zoneprobas} is then devoted to the motivation and exact definition of the model. In particular we shall define a clearance zone and the probabilities associated to the zone being empty.
Exact and phenomenological analytic results are then presented in section \ref{section:analytic} and numerical ones in section \ref{section:numeric}. Section \ref{section:ccl} concludes the paper.

\section{TASEP with a step initial condition}
\label{section:introstep}

\subsection{The model}
\label{subsection:defmodel}

The TASEP is defined on a one-dimensional lattice, whose sites may be \textit{occupied} or \textit{not} by a particle. During time evolution particles are allowed to hop from the site they occupy towards the site directly to the right, say from site $i$ to $i+1$, provided this latter site is \textit{empty} (not occupied). The system therefore satisfies the \textit{simple exclusion} constraint, \textit{i.e.} there is a maximal number of one particle on each site at any time. In this work we choose to evolve the system in \textit{continuous time}, where all the allowed transitions occur with the same constant rate $p$. Equivalently, waiting times between hopping attempts are drawn from an exponential distribution with parameter $p$.

\begin{figure}
\begin{center}
\scalebox{0.9}
{\includegraphics{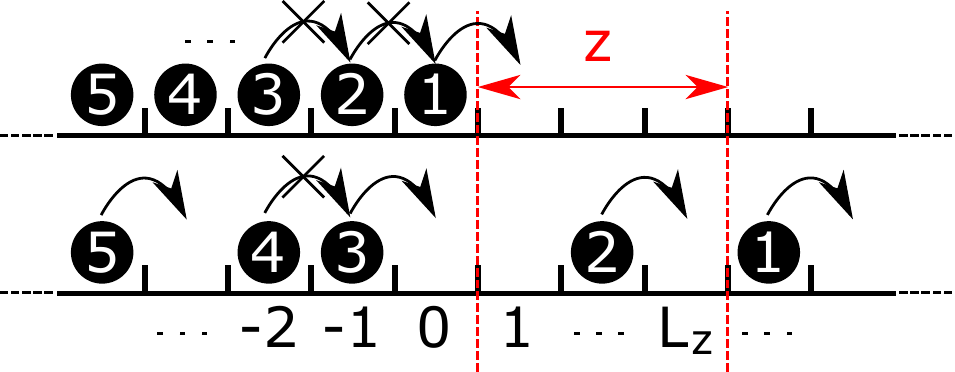}} 
\end{center}
\caption{\small Schematic representation of the system at initial time (top line) and at some arbitrary later time (bottom line), for $\nz = 3$. Particles are represented by black disks and allowed transitions are figured by arrows. The clearance zone $\Z$, defined in section \ref{section:zoneprobas}, is a zone of size $\nz$ located to the right of the initial discontinuity. }  
\label{fig:scheme}
\end{figure}

Here we will consider the TASEP on an infinite line.
We shall study the relaxation of a \textit{step} initial condition, see Fig.\,\ref{fig:scheme}.
At initial time, 
all the sites to the left of a given site are occupied and all the sites to the right are empty. 

Sites are labeled by integers ranging from $-\infty$ to $\infty$. We write $\tau_i(t) = 0$ if site $i$ is empty and $\tau_i(t)=1$ if site $i$ is occupied at time $t$. 
Setting the discontinuity between sites $0$ and $1$, the initial condition reads
\begin{equation}
 \label{eq:icni}
\tau_i(0) = 
\left\{
\begin{array}{c l}     
    1 & i \leq 0\\
    0 & i > 0
\end{array}\right.
\end{equation}
Equivalently, the state of the system may be described by the positions of the particles that occupy it. More precisely, we label the particles with a positive integer $k=1,2,\ldots$ from right to left.
The state of the system at any time $t$ is therefore described by a formally infinite vector $\bx = (x_1(t),x_2(t),\ldots)$ and the initial condition reads
\begin{equation}
 \label{eq:icxk}
x_k(0)=x_k^0 \equiv-k+1 \;\;\; \forall k \ge 1.
\end{equation}
Notice that the positions of the particles are ordered at all times, \textit{i.e.} $x_1(t) > x_2(t) > \ldots$ for all $t$. Here the only rate in the system is $p$, and time may be rescaled to set $p=1$, which we shall do in the following.

Numerically, the system is simulated by evolving the vector of the positions of the first $N$ particles $\bx_N(t)$ in continuous time using Gillespie's algorithm~\cite{gillespie1977}.	As the motion of one particle is not influenced by the particles on its left, the numerical measurement of any observable involving only the first $N$ particles will be free of finite size effects. For other observables for which the number of particles which are involved is not known a priori, one must check that enough particles
	are simulated. It is possible to verify during the simulations that the leftmost simulated particle does not move during any of the realizations of the Monte Carlo simulation, to ensure that the measurements are free of finite-size effects. In practice, simulating a step initially from $-100$ to $-1$ gives negligible finite-size effects for all results presented in this paper.

In the next subsection we review some theoretical results on the TASEP that we will use in the following.

\subsection{Propagator}
\label{subsection:prop}

The propagator of the infinite TASEP with a finite arbitrary number of particles $N$ has been obtained in Ref.\,\cite{schutz1997} by solving the master equation. The dynamics is obviously invariant under time translations, so that the propagator depends only on the difference $t$ between the initial and the final time. The probability to reach configuration $\bx$ starting from $\by$ then reads
\begin{equation}
 \label{eq:prop}
G_N (\bx,\by;t) = \det [F_{l-j} (x_{N-l+1}-y_{N-j+1};t)]_{j,l = 1,\ldots,N},
\end{equation}
where we keep track of the number of particles~$N$ and the functions $F_n(x;t)$ will be defined shortly. We also stress the fact that the propagator~(\ref{eq:prop}) exactly describes the motion of particles $1$ to $N$ since the motion of particles with $k \geq N+1$ has no effect on the $N$ first particles. It is manifest in~(\ref{eq:prop}) that the system is invariant under spatial translations.

The $F_n(x;t)$ have the following expression~\cite{schutz1997, sasamoto2007}
\begin{equation}
 \label{eq:Fn}
F_n(x;t) \equiv \frac{1}{2 \pi \ci} \oint_{\Gamma_{0,-1}} \frac{w^{-n}}{(1+w)^{x+1-n}} \ee^{w t} \dd w,
\end{equation}
where $\ci^2 = -1$ and $\Gamma_{0,-1}$ is a contour that encloses both potential singularities $0$ and $-1$. It is clear from~(\ref{eq:Fn}) that $F_n(x;t) = 0$ for $n \leq 0$ and $x \leq n-1$. The $F_n(x;t)$ have several remarkable properties listed in Ref.\,\cite{schutz1997}. In the following, we will in particular use
\begin{equation}
 \label{eq:sumFn}
\sum_{z=x}^\infty F_n(z;t) = F_{n+1}(x;t).
\end{equation}
Simple arguments show that at long times $F_n(x;t) \sim t^{n-1}$ if $n > 0$ and $F_n(x;t) \sim t^{x-n} \ee^{-t}$ for $n \leq 0$ and $x > n-1$. Finally, the first few $F_n(x;t)$, which we will need later, read
\begin{eqnarray}
 \label{eq:Fnsmall}
 F_{-1}(x;t) &= \frac{t^{x}}{(x+1)!} (x+1-t) \ee^{-t} \Theta(x +1 \geq 0), \nonumber \\
 F_0(x;t) &= \frac{t^x}{x!} \ee^{-t} \Theta(x \geq 0), \nonumber \\
 F_1(x;t) &= 1- \ee^{-t} \sum_{k=0}^{x-1} \frac{t^k}{k!}, & \\
 F_2(x;t) &= 1+t-x - \ee^{-t} \sum_{k=0}^{x-2} \frac{t^k}{k!} (x-k-1), &\nonumber 
\end{eqnarray}
where $\Theta(a)=1$ if assertion $a$ is true and $0$ otherwise, and it is understood that a sum with ill-ordered bounds vanishes.
In particular, the number of hops of an isolated particle follows a Poisson law $G_1(x,y;t) = F_0(x-y;t) = \frac{t^{x-y}}{(x-y)!} \ee^{-t}$ for $x-y \geq 0$.

As the dynamics of a TASEP particle is not influenced by its successor, the results presented here for finite $N$ may as well be applied to a system with an infinite number of particles.
More precisely, starting from a semi-infinite step at time $t=0$, the probability to find the $N$ rightmost particles at positions $\bx$ at time $t$ is $G_N(\bx,\bx^0;t)$, where $\bx^0$ is defined by~(\ref{eq:icxk}).

From there, the expression of the probability that the $N^\mathrm{th}$ particle has hopped at least $M$ times by time $t$ was found to be~\cite{nagao_s2004}
\begin{eqnarray}
 \label{eq:probxNhopM}
&& \Pr[ x_N(t)-x^0_N \geq M]\nonumber\\ &=& \sum_{x_1 > x_2 > \ldots > x_N \geq x_N^0 + M} G_N(\bx,\bx^0;t) \nonumber\\
&=& \sum_{x_1 \geq x_1^0 + M} \sum_{x_2 \geq x_2^0 + M} \ldots \sum_{x_N \geq x_N^0 + M} G_N(\bx,\bx^0;t)\\
&=& \det [F_{l-j+1} (x^0_{N-l+1}+M-x^0_{N-j+1};t)]_{j,l = 1,\ldots,N} \nonumber \\
&\equiv& G^+_N(\bx^0 +M,\bx^0;t).\nonumber 
\end{eqnarray}
The first equality comes from the definition of the propagator, the second one is the technical step, in which the sums are decoupled using~(\ref{eq:sumFn}) and the antisymmetry of the determinant. At the third one the summations are performed columnwise using~(\ref{eq:sumFn}) again, and the fourth one is simply a shorthand that stresses the fact that all the indices of the $F_n(x;t)$ have simply been increased by $1$ compared to the propagator~(\ref{eq:prop}). More details on the calculation can be found in Refs.\,\cite{nagao_s2004, sasamoto2007}.

The previous results give information about the correlations between sites at
a given time. There are also results
about some other types of correlations, for example
for the positions of a given particle at different times~\cite{johansson2016}.
However, the general correlation
for any times/positions is not known.

\subsection{Hydrodynamic profile and maximal current}
\label{subsection:hydro}

The properties of the system at long times have been investigated using the
hydrodynamic approximation. 
At scales very large compared to the lattice spacing, the variable $i$ can be replaced by a continuous
	space variable $x \in (-\infty,\infty)$.
In order to describe the system at the large scales, one 
defines the density field $\rho(x,t)$.
In the case of the TASEP,
at large times the evolution of the density profile is described to
lowest order by 
Burgers' equation~\cite{spohn1991,benarous_c2011}.
\begin{equation}
 \label{eq:burgersinvisc}
\frac{\partial \rho}{\partial t} = - (1-2\rho) \frac{\partial \rho}{\partial x},
\end{equation}
with the initial condition $\rho(x,0) = \Theta(x \leq 0)$.
The solution of the Burgers' equation then reads
\begin{equation}
\rho(x,t) = \left\{ \begin{array}{lr}
1 & x<-t\\
0 & x>t\\
(1-x)/2 & -t<x<t
\end{array}
\right.
\label{eq:solburgers}
\end{equation}

Actually there are fluctuations around this profile,
which have been characterized in~\cite{johansson2000,benarous_c2011}.
In the long time limit, around the position of the initial step, the density profile
will flatten and tend towards $1/2$, while the corresponding current will tend
towards $\frac{1}{4}$.

\vspace{5mm}

In the next sections our interest is to find the probability for a given zone
(that we shall call \textit{ clearance zone} in the following)
to be empty, possibly for the first time. This requires, firstly, to
characterize better the short time behavior and, secondly, to account for some 
time correlations. Though the results of subsection~\ref{subsection:prop} are
valid at all times, the focus of most of the applications of these results was
on the long time behavior~\cite{sasamoto2007}. Besides, we are missing some
information on time correlations. 

\section{Clearance zone}
\label{section:zoneprobas}

\subsection{Definition of the clearance zone}
\label{subsection:defzone}

Problems in which a certain set of sites may or may not be accessible to the incoming particles arise in some applied TASEP-based models. In Ref.\,\cite{turci2013} Turci \textit{et al.} studied a model for mRNA in which a zone could open or close at certain rates. In their model the lattice represents a mRNA strand and the particles represent ribosoms, which synthetize proteins by moving along the mRNA strand. The motion of the ribosoms may be blocked by some buckling of the mRNA or by other proteins that may attach on a specific region of the mRNA strand if this region is empty. This feature is accounted for in the model by defining a zone that corresponds to this specific region in the TASEP lattice, and can flip between open and closed states. As in the biological system, the blocking may occur only if the region is empty.

Jelic \textit{et al.} studied a model of pedestrians in which two counter-propagating flows of pedestrians share a common bottleneck of several sites in which only one species can enter at a time~\cite{jelic_a_s2012}. The other species is then prevented to enter the bottleneck until the bottleneck-zone is empty.

In both examples, if the parameters are such that the clearance zone stays closed for a long enough time, impeded particles tend to accumulate behind the zone. The flow when the zone opens again is then very similar to the relaxation of a step profile.

We therefore believe that it is of interest to 
study a minimal version of
this problem. For an infinite TASEP with a step initial
condition as detailed in subsection~\ref{subsection:defmodel}, we define the
\textit{clearance zone} $\Z$ as the set of sites $1$,$2$,\ldots,$\nz$. The size
$\nz$ of the clearance zone is the only parameter in the problem. Such a
clearance zone is represented on Fig.\,\ref{fig:scheme}. It can be thought as
being closed until the initial time $t=0$.

\subsection{Probabilities to be empty}
\label{subsection:defprobas}

\begin{figure}
\begin{center}
\scalebox{0.5}
{\includegraphics{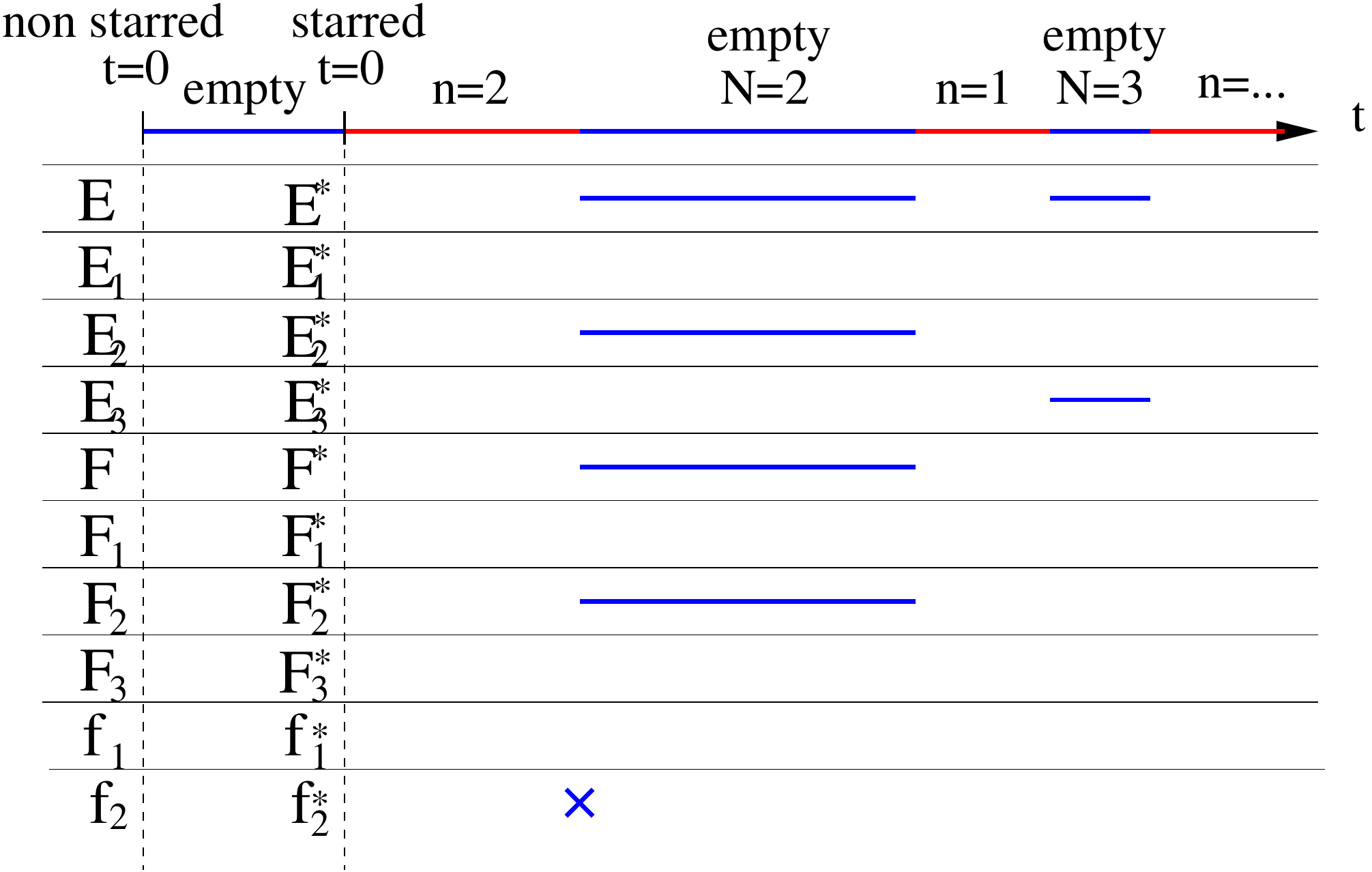}} 
\end{center}
\caption{\small Time intervals contributing to the studied probabilities for a sample realization of the dynamics. The axis of time is the horizontal axis at the top of the picture. In a particular history the state of the clearance zone $\Z$ alternates between 'empty' (blue intervals) and 'occupied' (red intervals), and the zone $\Z$ is empty at $t=0$. During each of the occupied intervals a certain number of particles $n$ goes through the clearance zone and exits it, \textit{i.e.} hops from $\nz$ to $\nz+1$. The total number of particles that has exited up to time $t$ is $N$, the sum of all $n$ until then. Under the time axis the studied probabilities are represented, and the intervals that contribute to them are reported on their line. The starred quantities record the same intervals as their non-starred counterparts, but their initial time is defined to be the moment where the zeroth empty interval ends.}  
\label{fig:defprobas}
\end{figure}

The first quantity of interest to us will be the probability that the clearance zone $\Z$ is empty, \textit{i.e.} that $\tau_1(t) = \tau_2(t) = \ldots = \tau_\nz(t) = 0$ at some given time $t$. We will also be interested in the number of particles to the right of $\Z$ when the clearance zone is empty. All the subsequent definitions are illustrated in Fig.\,\ref{fig:defprobas}. 

More precisely, we define $\cE_N(t)$ as the probability that $\Z$ is empty at $t$ and that $N$ particles have passed through the clearance zone. In a more mathematical way
\begin{eqnarray}
 \label{eq:defEN}
\cE_N(t) &\equiv& \Pr\Bigg[\prod_{k=1}^\nz (1-\tau_k(t)) = 1 \;\&\; \sum_{k=\nz+1}^\infty \tau_k(t) = N\Bigg] \nonumber \\
&=&  \Pr[x_N(t) > \nz \;\&\; x_{N+1}(t) \leq 0].
\end{eqnarray}
We also define $\cE(t) = \sum_{N=1}^\infty \cE_N(t)$, the total probability that the clearance zone is empty at time $t$. 

Note that the sum defining $\cE(t)$ starts with $N=1$, \textit{i.e.} one considers only the case of an empty zone after the passage of at least one particle through the clearance zone. There is actually a time interval starting at $t=0$ where $\Z$ is empty and no particle has gone through it. In the future we may want to change the time origin in order to eliminate this 'zeroth' empty interval. We shall therefore consider an alternative definition of the initial time and define the $\cE^*_N(t)$ and $\cE^*(t)$ analogously to $\cE_N(t)$ and $\cE(t)$, except that the initial time of the starred quantities will be taken at the end of the zeroth empty interval. The zeroth empty interval ends when particle $1$ enters the clearance zone by performing its first hop from site $0$ to site $1$. The waiting time before the first hop is exponentially distributed, and the starred and non-starred quantities are related by
\begin{equation}
 \label{eq:phiphist}
\phi(t) = \int_{s=0}^t \ee^{-(t-s)} \phi^*(s) \dd s,
\end{equation}
for $\phi = \cE_N,\cE$ and similar functions to be defined later (which will all verify $\phi(0) = 0$, a condition for~(\ref{eq:phiphist}) to hold) . Eq.\,(\ref{eq:phiphist}) may be inverted to give
\begin{equation}
 \label{eq:phistphi}
\phi^*(t) = \phi(t) + \frac{\dd \phi}{\dd t}(t).
\end{equation}
Equivalently, the initial condition for the starred quantities is 
\begin{equation}
 \label{eq:icstxk}
x_k(0)=x_k^{0*} \equiv
\left\{
\begin{array}{c l}     
    1 & k=1\\
    -k+1 & k > 1
\end{array}\right.
\end{equation}

In some models the clearance zone is allowed to close when it is empty, so that histories where there exist empty intervals will have modified weights compared to what the model studied here would give. As a first step to account for it, we shall also be interested in the first time $\Z$ is empty (not considering the zeroth one), though no closure will be considered in the present paper. We define $\cF_N(t)$ as the probability that $\Z$ is empty at time $t$, conditioned by the fact that $t$ belongs to the first time interval where $\Z$ is empty, and that $N$ particles have passed. We also define $\cF(t)$, the $\cF_N^*(t)$ and $\cF^*(t)$ in analogy with the $\cE$ quantities. We have $\cF_N(t) \leq \cE_N(t)$, with the special case $\cF_1(t) = \cE_1(t)$.

Finally we also define the $f_N(t) \dd t$, $f(t) \dd t$ and their starred counterparts as the probabilities that $\Z$ becomes empty for the first time between $t$ and $t+\dd t$ with $N$ particles passed. The clearance zone almost surely gets empty for the first time at some moment of time, which imposes
\begin{equation}
 \label{eq:normfN}
 \int_{t=0}^\infty f(t) \dd t = 1.
\end{equation}
Eq.\,(\ref{eq:normfN}) is true when replacing $f$ by $f^*$ as well.

A quantity similar to $f(t)$ has been introduced in
Ref.\,\cite{gabrielli_t_v2013} and extended in Ref.\,\cite{barre_t_v2013,
talbot_g_v2015}, in a model in continuous space,
where particles are injected at the entrance
of a one-dimensional channel
and cross it with a constant velocity until
they reach the exit.
In their model clogging occurs if more than one particle
is present at the same time in the channel, and the authors derive various exact formulas for
reversible~\cite{gabrielli_t_v2013} or irreversible~\cite{barre_t_v2013}
clogging, including the probability that no clogging occurs until a given time
in the irreversible case. Adapting this simple Markovian model to our case
would give an exponential distribution for $f^*(t)$.

The characterization of the $\cE$ and $\cF$ quantities will be the main goal of this paper. Obviously the knowledge of a non-starred quantity is equivalent to the knowledge of its starred counterpart, so that we shall use the most convenient set of functions, depending on the situation. It is also readily noticed that the $\cE_N$ quantities are single-time correlation functions and are expected to be a lot easier to compute than the $\cF_N$ quantities, which depend on the whole history of the system.

In the following we present a combination of exact, approximate and numerical arguments that account for the main properties of the probabilities defined in this subsection.

\section{Analytic results}
\label{section:analytic}

\subsection{Direct calculation for $N=1$}
\label{subsection:combNe1}

Here we focus on the simplest case $N=1$, where all the quantities of interest can be computed directly, for an arbitrary length $\nz$ of the clearance zone $\Z$.
In this section we shall show how to compute $f^*_1(t)$, while the expressions for $\cE_1(t) = \cF_1(t)$ and $\cE_1^*(t) = \cF_1^*(t)$, that can be obtained analogously, will be given in~\ref{section:exprF1}.
By definition of $f^*_1(t)$ we track the events where $x_1(t) > \nz$ and $x_2(t) \leq 0$ and calculate their associated probability. By definition of the starred quantities, particle $1$ is on site $1$ at $t=0$. Since $N=1$ and the motion of the second particle is not influenced by particles $k > 2$, we may restrict ourselves to a $2$-particle system with $x_1(0)=1$ and $x_2(0)=-1$. We define two remarkable times: the time when particle $1$ hops from $1$ to $2$ will be denoted $s$ and the time when particle $2$ hops from $-1$ to $0$ (if it does) will be denoted $s'$. For the history to contribute to $f_1^*(t)$, we must obviously have $s < t$. 

We shall now compute the probabilities of the histories that will contribute to $f_1^*(t)$.
We first consider the case $s' < s$. The probability to have a given value of $s$ and any value $s' < s$ then reads $\ee^{-s} \dd s \int_{s'=0}^s \ee^{-s'} \dd s' = \ee^{-s} (1-\ee^{-s})  \dd s$. In that case particle $2$ is necessarily blocked on site $0$ until $1$ hops at time $s$. After that, the events that contribute to $f_1^*(t)$ are those where particle $2$ does not hop during an interval of length $t-s$, with associated probability $\ee^{-(t-s)}$, while particle $1$ must have hopped for the $(\nz-1)^\mathrm{th}$ time at time exactly $t$, which occurs with probability $\frac{(t-s)^{\nz-2}}{(\nz-2)!} \ee^{-(t-s)} \dd t$. 

In the second case $s' > s$, the probability to have a given value of $s$ and a value of $s'$ such that $s' > s$ is $\ee^{-2 s} \dd s$. Particle~$2$ is then allowed to perform at most one hop between $s$ and $t$, corresponding to a probability $\ee^{-(t-s)} + (t-s) \ee^{-(t-s)}$ to perform respectively $0$ or $1$ hop. 

The sum of these two terms reads
\begin{eqnarray}
\label{eq:cfst1comb}
 f^*_1(t) &=&\int_{s=0}^t \dd s \ee^{-s} (1-\ee^{-s}) \frac{(t-s)^{\nz-2}}{(\nz-2)!} \ee^{-2 (t-s)}\nonumber \\&&+ \int_{s=0}^t \dd s \ee^{-2s} \frac{(t-s)^{\nz-2}}{(\nz-2)!} \ee^{-2 (t-s)} (1+t-s) \\
&=& \frac{ (\nz-1)}{\nz!} t^{\nz} \ee^{-2 t}+\bigg(1-\frac{\Gamma(\nz-1,t)}{(\nz-2)!}\bigg) \ee^{-t}, \nonumber
\end{eqnarray}
where $\Gamma(n,t) \equiv \int_{s=t}^\infty s^{n-1} \ee^{-s} \dd s$ is the upper incomplete gamma function and $\Gamma(n) \equiv \Gamma(n,0)$ is Euler's gamma function. 

From Eq.\,(\ref{eq:cfst1comb}) we see that $f^*_1(t)$ decays as $\ee^{-t}$ at large times. 
In a naive reasoning we could consider that the events contributing to the long time behaviour of $f^*_1(t)$ are those where particle $1$ goes very slowly through the zone $\Z$ and particle $2$ does not enter the clearance zone.
However the impediment of two particles would give a contribution $\ee^{-2 t}$.
Actually, the dominant contribution is given by events where particle
$1$ blocks particle $2$ most of the time. This requires the blocking of only
one particle, hence a probability~$\sim \ee^{-t}$.

This combinatorial method provides an intuitive way of computing $f^*_1(t)$. It however seems hard to extend to $N > 1$. This shall be done in the next subsection by using the exact expression of the propagator.

\subsection{The determinantal formula}
\label{subsection:det}

Here we will see that an exact formula can be found for the $\cE_N(t)$ on the basis of the exact expression for the propagator~(\ref{eq:prop}) and on the result~(\ref{eq:probxNhopM}).

For determining $\cE_N(t)$ for a given $N$ one only needs to consider the $N+1$ first particles. We have
\begin{eqnarray}
 \label{eq:ENdet1}
\cE_N(t) &=& \Pr[x_N(t) > \nz \;\&\; x_{N+1}(t) \leq 0] \nonumber\\ &=&\sum_{i = x_{N+1}^0}^{0} \Pr[x_N(t) > \nz \;\&\; x_{N+1}(t) = i] \\ &=&\sum_{x_{N+1} = x_{N+1}^0}^{0} \sum_{x_1 > x_2 > \ldots > x_N > \nz} G_{N+1}(\bx,\bx^0;t). \nonumber
\end{eqnarray}
By a trick similar to the one used in equation~(\ref{eq:probxNhopM}), the summations over the $x_{k}$ with $k \leq N$ may be decoupled and carried out, resulting in an increase of the index of the $F_n(x;t)$ in $N$ columns of the matrix. We get as an intermediate result
\begin{equation}
 \label{eq:ENdet2}
\Pr[x_N(t) > \nz \;\&\; x_{N+1}(t) = i] = \det A_{jl} \big|_{j,l=1,\ldots,N+1}
\end{equation}
with
\begin{equation}
 \label{eq:Ajl}
A_{jl} \equiv
\left\{
\begin{array}{c l}     
    F_{1-j}(i-x^0_{N-j+2};t) & l=1\\
    F_{1+l-j}(\nz+1-x^0_N+x^0_{N-l+2}-x^0_{N-j+2};t) & l > 1
\end{array}\right..
\end{equation}
The sum over $i$ only affects the first column and can be carried out, using~(\ref{eq:sumFn}) once again. We finally get
\begin{equation}
 \label{eq:ENdetfin}
\cE_N(t) = \det E_{jl} \big|_{j,l=1,\ldots,N+1}
\end{equation}
with
\begin{equation}
 \label{eq:Ejl}
E_{jl} \equiv
\left\{
\begin{array}{c l}     
    F_{2-j}(x^0_{N+1}-x^0_{N-j+2};t) - F_{2-j}(1-x^0_{N-j+2};t) & l=1\\
    F_{1+l-j}(\nz+1-x^0_N+x^0_{N-l+2}-x^0_{N-j+2};t) & l > 1
\end{array}\right..
\end{equation}

\begin{figure}
\begin{center}
\scalebox{0.4}
{\includegraphics{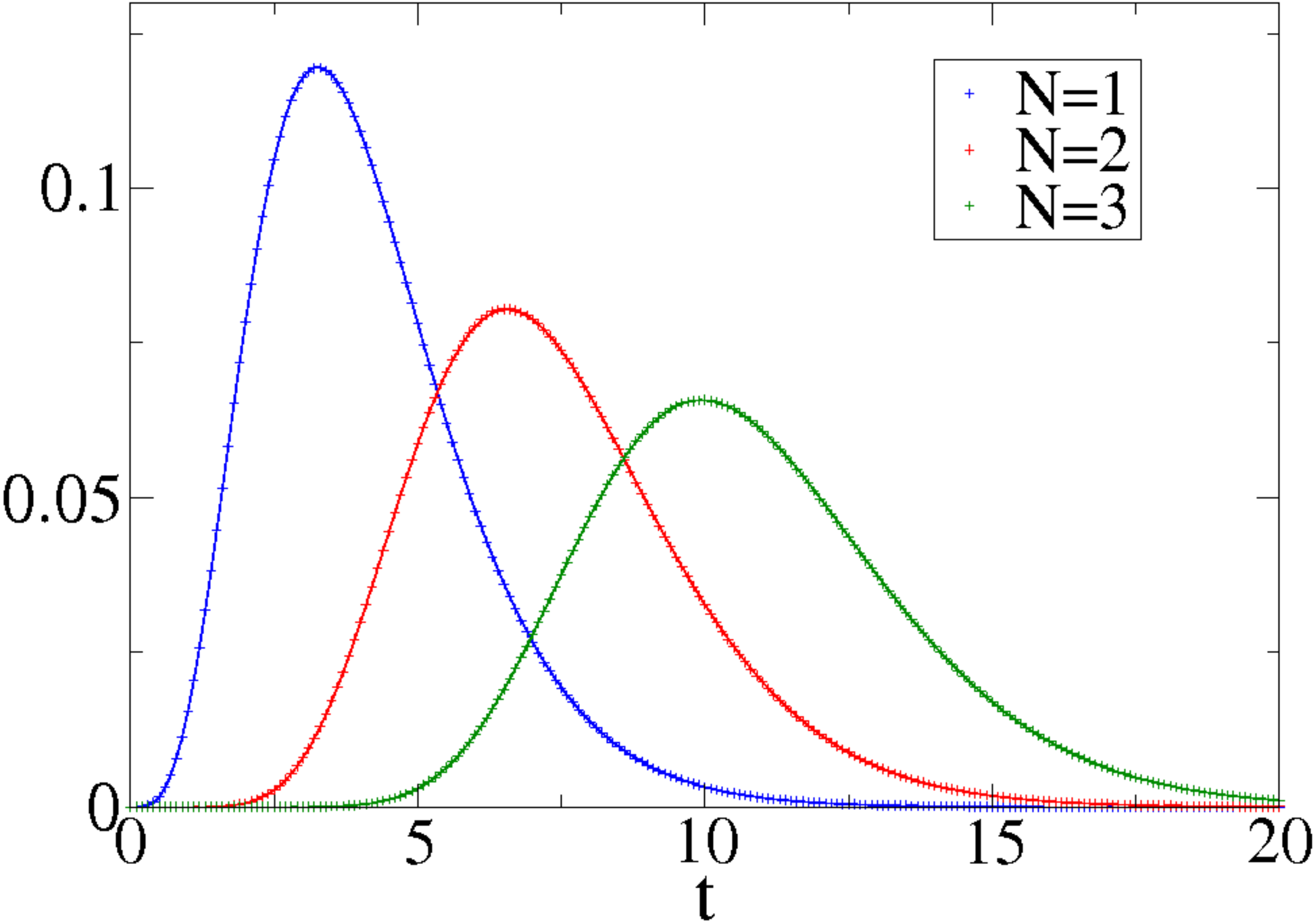}} 
\end{center}
\caption{\small Plot of the exact expression of $\cE_N(t)$ (solid lines) compared to numerics (pluses) for $\nz = 3$ and $N=1,2,3$.}  
\label{fig:ENlow}
\end{figure}

The result~(\ref{eq:ENdetfin})-(\ref{eq:Ejl}) is exact, and can be checked to give Eq.\,(\ref{eq:cFst1comb}) for $N=1$. The first few $\cE_N(t)$ are plotted in Fig.\,\ref{fig:ENlow}. The agreement with the numerics is excellent as expected.

Eqs. (\ref{eq:ENdetfin}) and (\ref{eq:Ejl}) give a compact exact solution to part of our problem. These expressions however involve $(N+1) \times (N+1)$ determinants and are not very easy to manipulate. In particular, it is not clear how to find a compact expression for $\cE(t)$ starting from (\ref{eq:ENdetfin})-(\ref{eq:Ejl}).

\subsection{$\cF_2(t)$}
\label{subsection:F2}

It was already mentioned that the $\cF_N(t)$ are more complicated quantities than the $\cE_N(t)$ because they depend on the whole history of the system. In principle exact expressions could be written by summing over intermediate variables. Here we simply do it for $N=2$ to illustrate the idea. We also derive the asymptotics of $\cF_2(t)$ at large times.

\begin{figure}
\begin{center}
\scalebox{0.4}
{\includegraphics{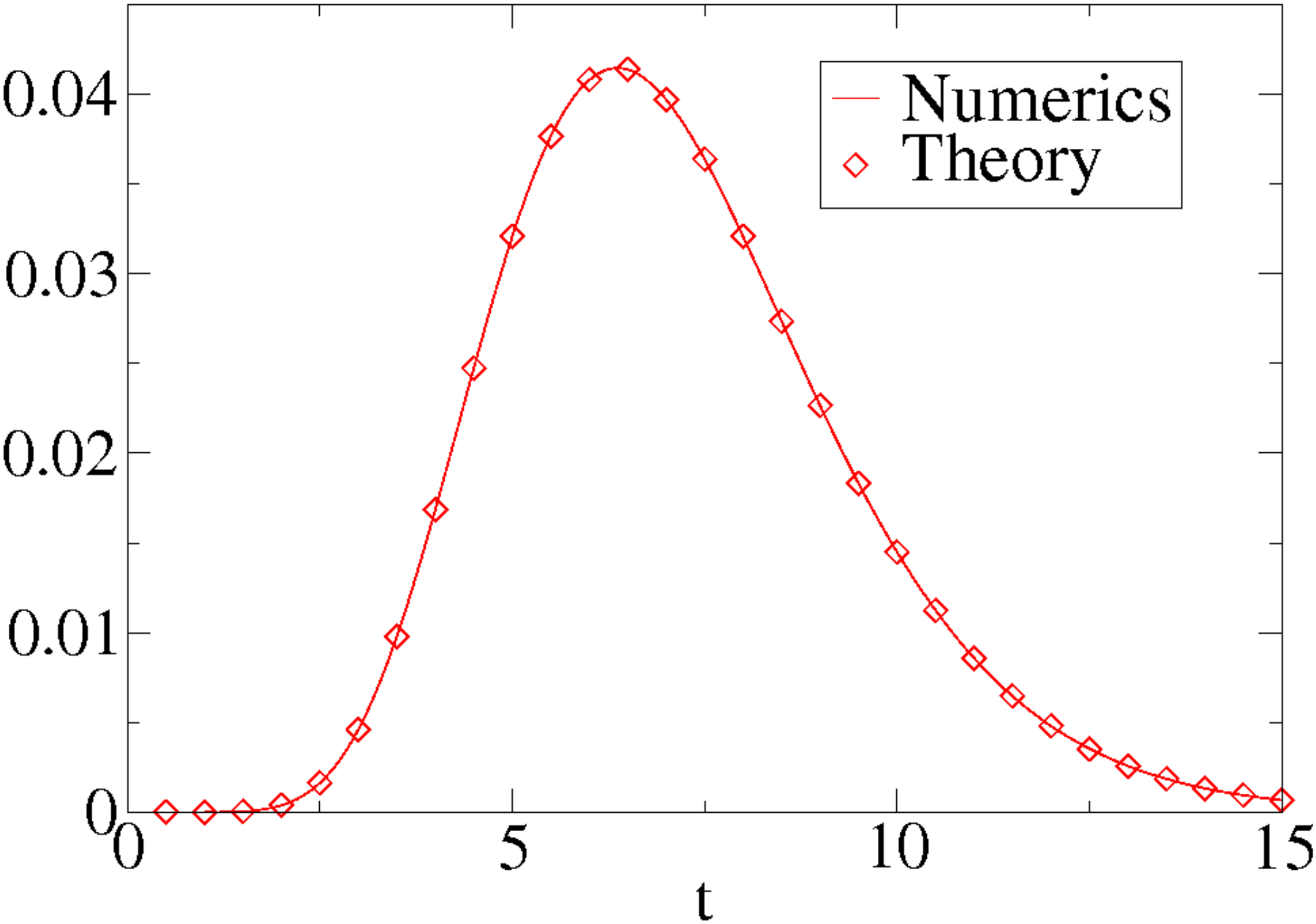}} 
\end{center}
\caption{\small Plot of the exact expression (\ref{eq:cF2det})
of $\cF_2(t)$ (red crosses) compared to numerics (black curve) for $\nz = 3$.}  
\label{fig:F2}
\end{figure}

To compute $\cF_2(t)$ we have to select events such that when the first particle exits $\Z$ the second already entered it.
We denote $s$ as the moment when particle $1$ exits the clearance zone,
and $\bx^s$ as the state just before the hop (state at time $s^-$).
We have to enforce the following constraints on $\bx^s$:
\begin{itemize}
	\item Particle $1$ is supposed to hop from $\nz$ to $\nz+1$ at time $s$, therefore $x^s_1 = \nz$.
 \item The clearance zone cannot be empty after particle $1$ hops. We therefore need $x^s_2 > 0$.
 \item The clearance zone must be empty after particle $2$ leaves it, therefore particle $3$ cannot enter it. Consequently we must have $x_3 \leq 0$ at all times and in particular $x^s_3 \leq 0$.
\end{itemize}
At time $s$ particle $1$ hops, with probability $\dd s$. Between time $s$ and time $t$ the second particle is supposed to exit $\Z$. Following the same logic, at time $t$ the positions of the particles $\bx^t$ satisfy the following constraints:
\begin{itemize}
 \item Particle $3$ has not yet entered the clearance zone, $x^t_3 \leq 0$.
 \item Particle $2$ has exited, $x^t_2 > \nz$, and particle $1$ is ahead of it $x^t_1 > \nz+1$.
\end{itemize}
All these constraints can be expressed by the formula
\begin{eqnarray}
\label{eq:cF2det}
\cF_2(t) &=&  \int_{s = 0}^{t} \dd s \sum_{x_2^s = 1}^{\nz - 1} \sum_{x_3^s = -2}^{0} G_3((\nz,x_2^s,x_3^s)|\bx^0;s) \nonumber \\
&\times& \sum_{x_1^t = \nz+2}^{\infty} \sum_{x_2^t = \nz+1}^{x_1^f -1} \sum_{x_3^t = x_3^1}^{0}  G_3(\bx^t|(\nz+1,x_2^s,x_3^s);t-s)
\end{eqnarray} 
where we have integrated over all the possible times $s$ when particle $1$ exits. Fig.\,\ref{fig:F2} shows the agreement between the exact expression~(\ref{eq:cF2det}) and numerical results.

Although this expression is complicated, its asymptotic behavior for large times may be extracted. In \ref{section:asymptF2} we show that at large times we have
\begin{eqnarray}
 \label{eq:cFdetasymp}
\cF_2(t) &=& \int_{s=0}^t \ee^{-(t-s)} (t-s) \ee^{-s} \frac{s^2}{2} \dd s (1+O(t^{-1})) \\
&=& \frac{t^4}{24} \ee^{-t} (1+O(t^{-1})). \nonumber
\end{eqnarray}
Eq.\,(\ref{eq:cFdetasymp}) provides us with an equivalent of $\cF_2(t)$ for long times. It can be shown that terms coming from sectors where one of $s$ and $t-s$ is finite are subdominant. This exponential part of the decay generalizes to higher $N$ values, as we expect $\cF_N(t)$ to be a convolution of polynomial factors times simple exponentials, which combine to finally give an overall $\ee^{-t}$. The power law in front of the exponential seems harder to predict for general $N$.

In the following subsection we try to find a more systematic, yet approximate method for computing the $\cF_N(t)$.

\subsection{Approximate recurrence formula}
\label{subsection:rec}

In this section, in order to calculate the $\cF_N(t)$, we shall make
an approximation that we expect to be all the more relevant
as the zone $\Z$ gets larger.
We first assume that the history of the system can be factorized as a product
of histories over the intervals where $\Z$ is empty and where $\Z$ is occupied,
and we assume that a step is reformed each time the clearance zone is empty.
The durations of empty intervals are therefore drawn from a distribution 
\begin{equation}
 \label{eq:defpin}
\pin(t) \equiv \ee^{-t},
\end{equation}
where $t$ is the time needed to occupy the clearance zone again. 

Based on the above assumptions, we may now write two series of identities that link the unknown functions $\cE_N$, $\cF_N$ and $f_N$. The first one reads
\begin{equation}
  \label{eq:exprFf}
\cF_N(t) = \int_{s=0}^t f_N(s) \int_{r=t}^\infty \pin(r-s) \dd s \dd r.
\end{equation}
Equation~(\ref{eq:exprFf})
means that the system is in the first empty interval at time $t$ if it has
been emptied at time $s$ for the first time and no particle entered until then,
\textit{i.e.} the time $r$ at which the next particle will enter the
clearance zone is larger than $t$. 

The second identity is typical of first-passage problems. It links the $\cE_N$, the $\cF_N$ and the $f_N$,
\begin{equation}
 \label{eq:recEFf}
\cE_N(t) = \cF_N(t) + \sum_{N'=1}^{N-1} \int_{s=0}^t f_{N'}(s) \cE_{N-N'}(t-s) \dd s.
\end{equation}
Eq.\,(\ref{eq:recEFf}) states that if $\Z$ is empty at time $t$, either it is
the first empty interval or the first empty interval started at some time $s$
and the zone $\Z$ has been occupied again since then.

The $f_N$ can be eliminated by using the starred quantities and using (\ref{eq:defpin}). After these transformations, (\ref{eq:exprFf})-(\ref{eq:recEFf}) imply
\begin{equation}
 \label{eq:recEstFst}
\cE^*_N(t) = \cF^*_N(t) + \sum_{N'=1}^{N-1} \int_{s=0}^t \cF^*_{N'}(s) \cE^*_{N-N'}(t-s) \dd s.
\end{equation}
We already know an explicit expression (\ref{eq:ENdetfin})-(\ref{eq:Ejl}) for
the $\cE_N(t)$. Equation (\ref{eq:recEstFst}) can therefore be used to compute
the $\cF_N^*(t)$. For the convolution in time we define the shorthand $(\phi * \psi) (t) = (\psi * \phi) (t) \equiv \int_{s=0}^t \phi(s) \psi(t-s) \dd s$. In
this notation Eq.\,(\ref{eq:recEstFst}) reads
$\cE^*_N = \cF^*_N + \sum_{N'=1}^N (\cF^*_{N'} * \cE^*_{N-N'})$.
Using $\cF_1^* = \cE_1^*$ one can express the $\cF^*_N$ as a
function of the $\cE^*_N$,
\begin{eqnarray}
 \label{eq:recFNsmall}
\cF^*_2 &=& \cE^*_2 - \cE^*_1 * \cE^*_1, \nonumber\\
\cF^*_3 &=& \cE^*_3 -2 \cE^*_2 * \cE^*_1 + \cE^*_1 * \cE^*_1 * \cE^*_1,\\
\cF^*_4 &=& \cE^*_4 -2 \cE^*_3 * \cE^*_1 - \cE^*_2 * \cE^*_2 + 3 \cE^*_2 * \cE^*_1 * \cE^*_1 - \cE^*_1 * \cE^*_1 * \cE^*_1 * \cE^*_1.\nonumber
\end{eqnarray}
Eq.\,(\ref{eq:recEstFst}) may be inverted for arbitrary $N$. If we parametrize the partitions of the integer~$N$ by a set of numbers $\{l_i\}$ such that the integer $i$ appears $l_i$ times in the partition, the $\{l_i\}$ satisfy $\sum_i i l_i = N$. With this notation the general formula reads
\begin{equation}
\label{eq:recFNgal}
 \cF^*_N = \sum_{\{l_i\} / \sum i l_i = N} {\sum_i l_i \choose l_1, l_2, \ldots} 
(-1)^{(\sum_i l_i) -1} \bigg[(\cE_1^*)^{l_1} * (\cE_2^*)^{l_2} * \ldots\bigg],
\end{equation}
where ${\sum_i l_i \choose l_1, l_2, \ldots}=\frac{(\sum_i l_i)!}{l_1! l_2! \ldots}$ is a multinomial coefficient.

  \begin{figure}
\begin{center}
    \includegraphics[width=0.48\textwidth]{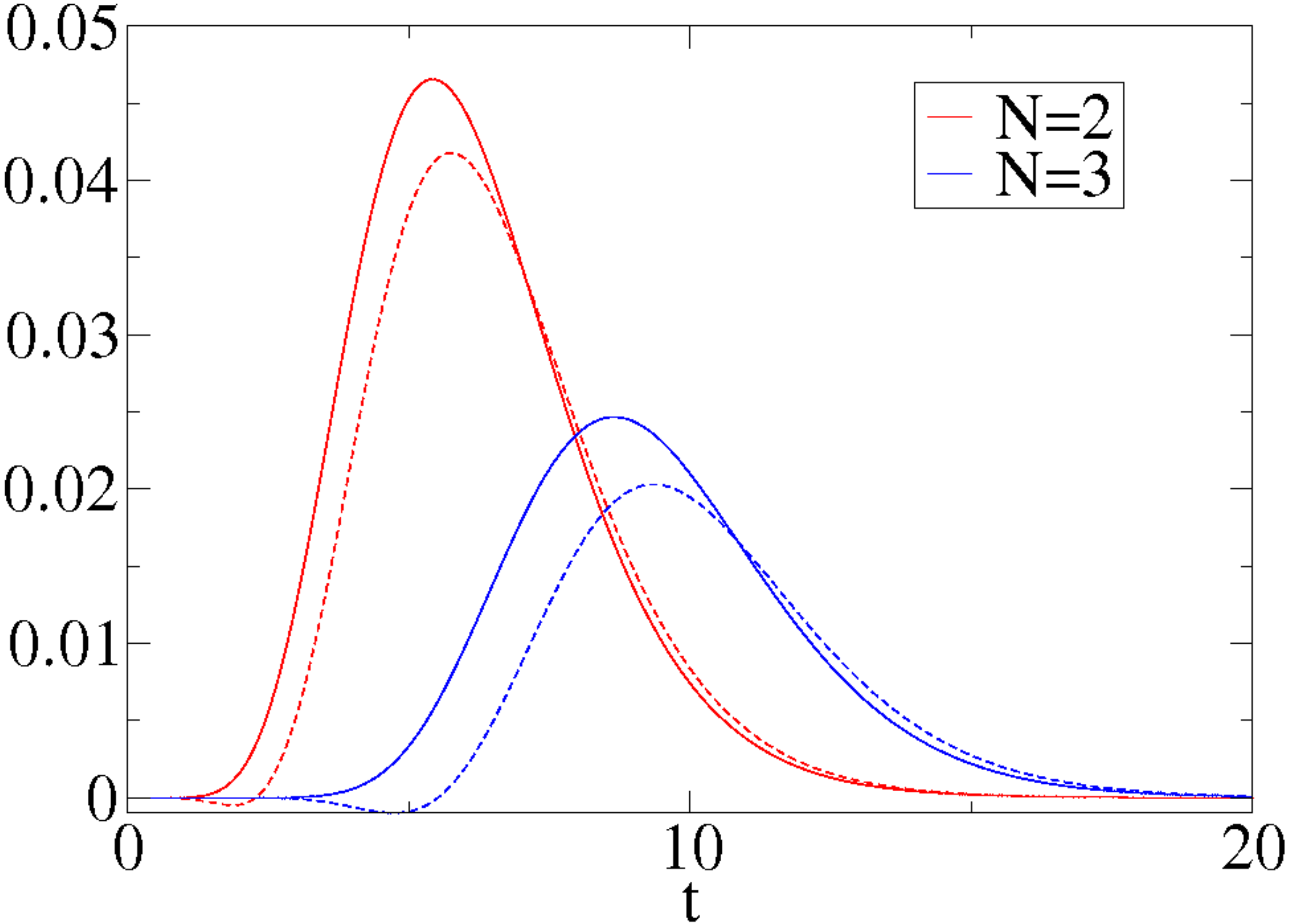} 
     \includegraphics[width=0.48\textwidth]{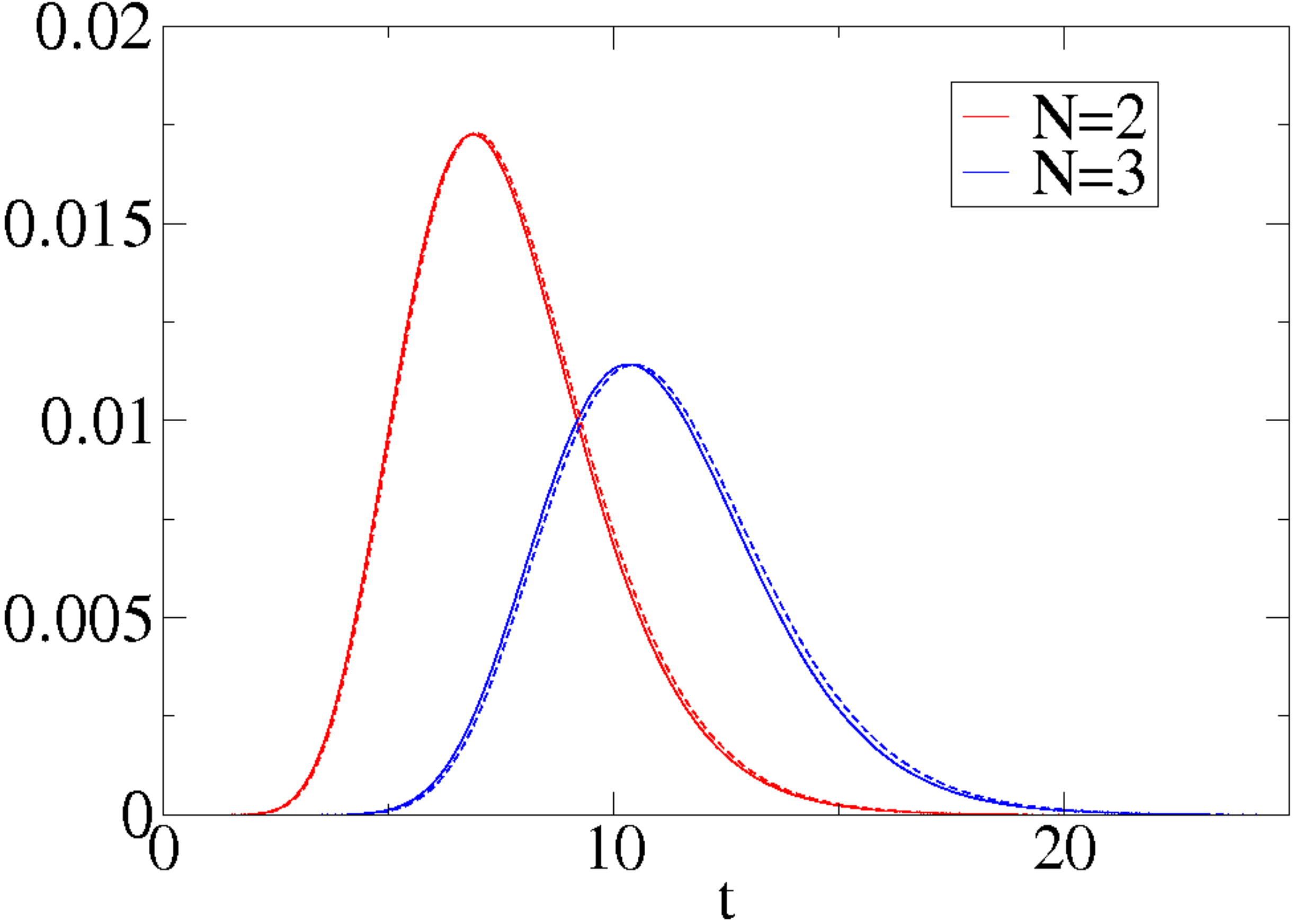} 
     \includegraphics[width=0.48\textwidth]{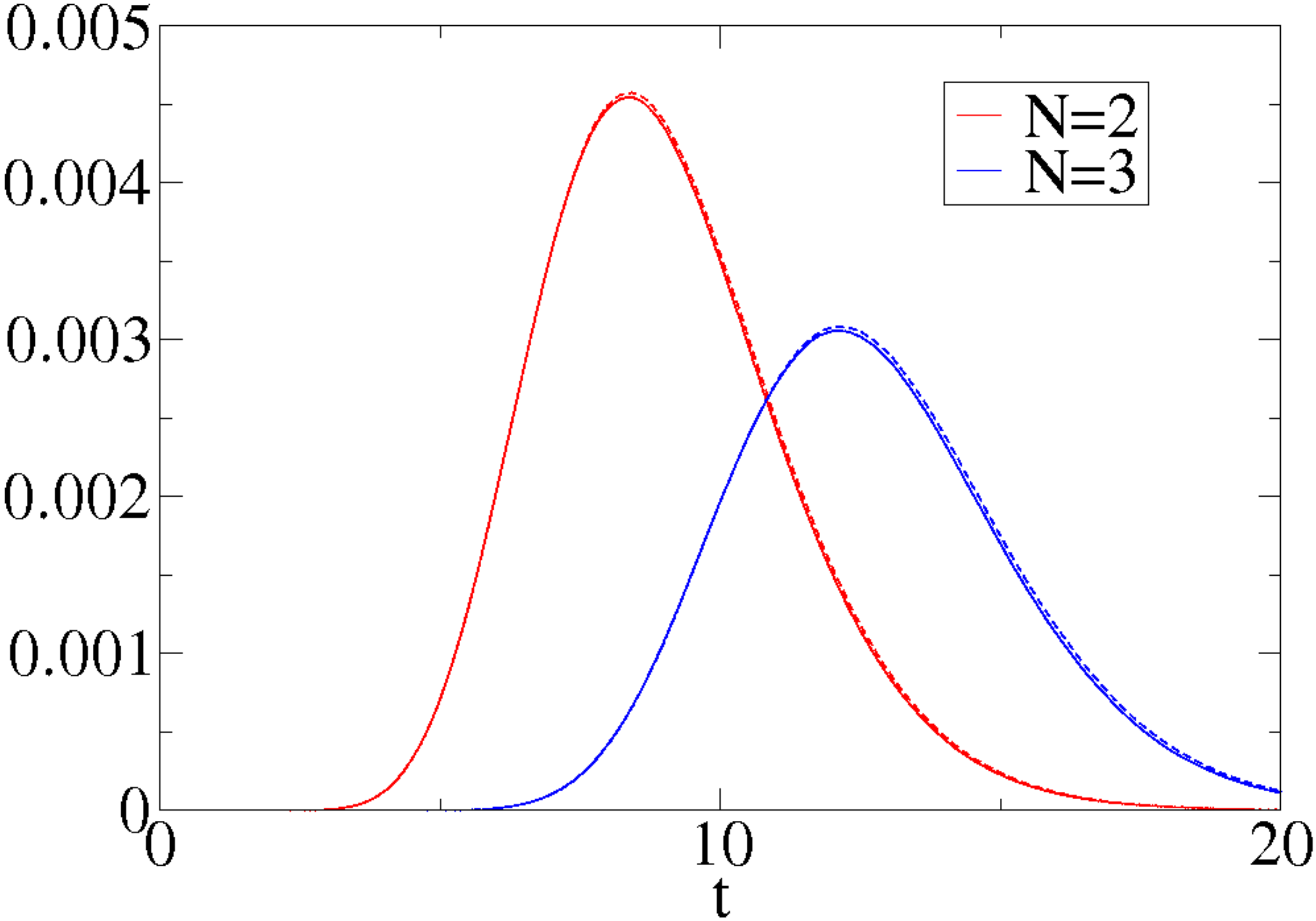} 
     \end{center}
\caption{\small Numerical tests of the convolution approximation~(\ref{eq:recFNsmall}) for $\cF^*_2(t)$ and $\cF^*_3(t)$ and for $\nz = 3$ (top right), $5$ (top left) and $7$ (bottom). Solid lines are direct numerical measurements of $\cF^*_2$ and $\cF^*_3$ while dashed lines are the RHSs of the first two equations~(\ref{eq:recFNsmall}).
\label{fig:testrec}
}
\end{figure}

The relations (\ref{eq:recFNsmall}) and (\ref{eq:recFNgal}) may be tested numerically. The results are shown in Fig.\,\ref{fig:testrec}, and it appears that the approximations (\ref{eq:exprFf}) and (\ref{eq:recEFf}) work better and better with increasing $\nz$. Indeed, histories where $\Z$ becomes empty after the $N^\mathrm{th}$ particle passes are most probably histories where the $(N+1)^\mathrm{th}$ particle has been blocked for a long time, hence an accumulation of particles behind. The larger the clearance zone, the longer the time particle $N+1$ has been blocked, the higher the density is likely to be on sites $i \leq 0$. For larger clearance zones the density profile is therefore closer to a step and the recurrence approximation indeed works better.

\section{Numerical results}
\label{section:numeric}

\subsection{Quantities summed over $N$}
\label{subsection:sumN}

A seemingly simpler problem concerns the summed quantities $\cE(t)$, $\cF(t)$ and $\cE^*(t)$, $\cF^*(t)$. We first note that, within the frame of the recurrence (\ref{eq:exprFf})-(\ref{eq:recEFf}), a simple relation exists between $\cE^*$ and $\cF^*$. Equation (\ref{eq:recEstFst}) can be summed over $N$ to give
\begin{equation}
\label{eq:recEtotFtot}
 \cE^* = \cF^* - \cE^* * \cF^*.
\end{equation}
We therefore focus on $\cE(t)$.

\begin{figure}
\begin{center}
\scalebox{0.4}
{\includegraphics{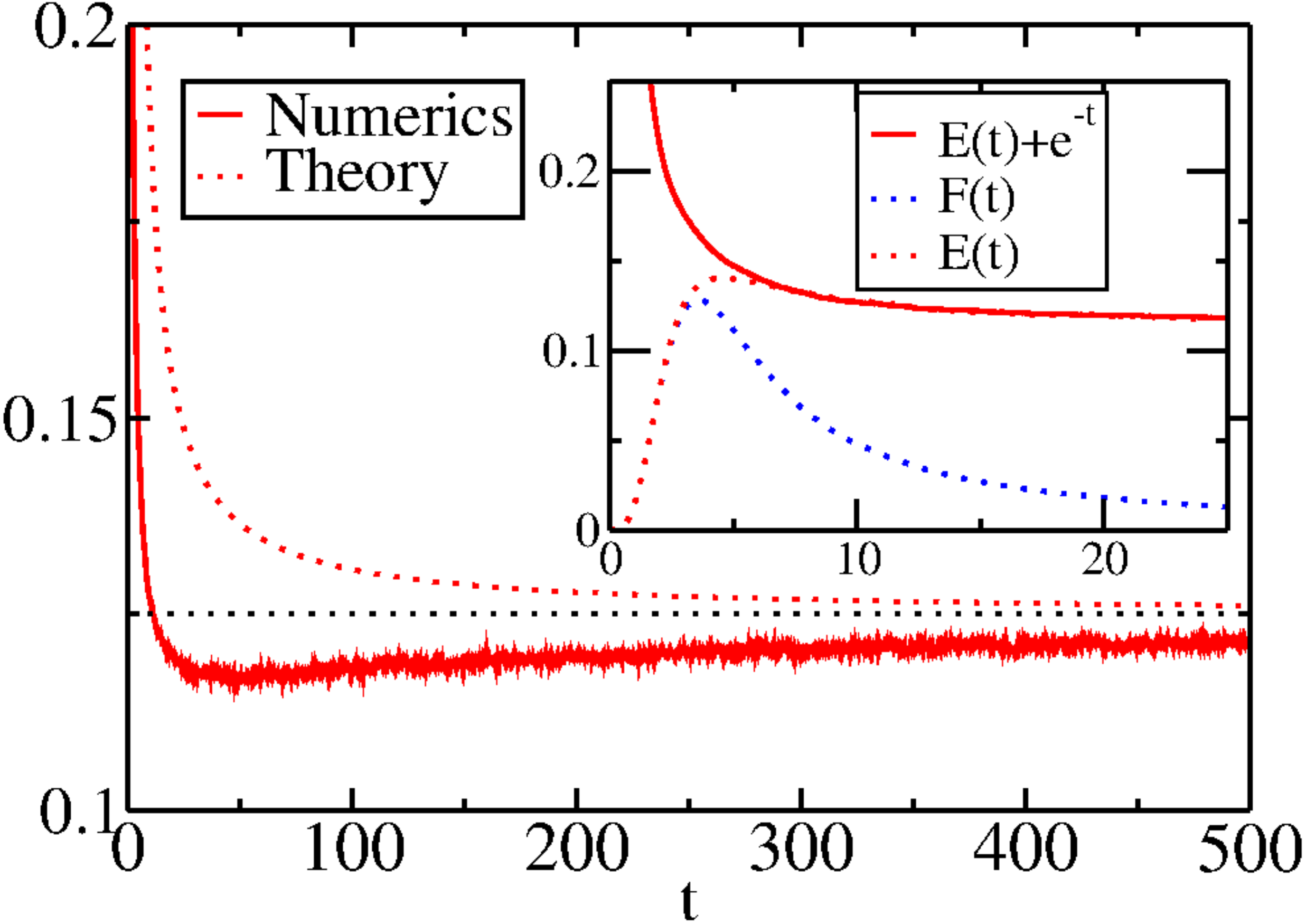}} 
\end{center}
\caption{\small Main graph: Numerical measurements of $\cE(t)+\ee^{-t}$ for $\nz = 3$ (red solid line). The black dashed line is constant and equal to the limit $2^{-3} = 0.125$ and the red dashed line is the prediction of~(\ref{eq:Etotproduct}) based on the hydrodynamic profile. Inset: Plot of $\cE(t)$ (red dashed), $\cE(t)+\ee^{-t}$ (red solid) and $\cF(t)$ (blue dashed line) at short times.}  
\label{fig:E}
\end{figure}

Although we have derived exact expressions for the $\cE_N(t)$, it is not clear how to sum over $N$ to obtain a useful expression for $\cE(t)$. For predicting $\cE(t)$ we try instead to make use of the hydrodynamic solution~(\ref{eq:solburgers}).
The initial step lies between sites $0$ and $1$, so that we expect
\begin{equation}
 \label{eq:Etotproduct}
 \cE(t) + \ee^{-t} \simeq \prod_{i=1}^{\nz} [1-\rho(i-0.5,t)],
\end{equation}
where the $\ee^{-t}$ on the LHS is the probability associated to the zeroth empty interval, which we added for convenience, though it is vanishing at hydrodynamic times. Eq.\,(\ref{eq:Etotproduct}) is compared to numerical measurements in Fig.\,\ref{fig:E}.

On Fig.\,\ref{fig:E} there is a discrepancy between the hydrodynamic
approximation and the numerics, which reveals that correlations exist.
The two curves have the same limit $2^{-\nz}$, corresponding to the convergence
towards a maximal-current phase where every site is occupied with probability
$1/2$.
The hydrodynamic prediction however converges from the top while the numerical
curve comes from the bottom. This effect can be attributed to the presence of
long-lived spatial correlations. Denoting ensemble averages with the brackets
$\langle \ldots \rangle$, it is numerically observed that at large times
two-point functions of the kind $\langle \tau_i (1-\tau_{i+1})\rangle (t)$ or
$\langle (1-\tau_i) \tau_{i+1} \rangle (t)$ are slightly larger than their stationary
value $1/4$, while those of the kind $\langle \tau_i \tau_{i+1}\rangle (t)$ and
$\langle (1-\tau_i) (1-\tau_{i+1})\rangle (t)$ are slightly smaller
than $1/4$. These correlations decay algebraically in time. As a
consequence, since neighbouring sites are anticorrelated we have $\cE(t)
< 2^{-\nz}$ at large times.

For future applications to closing regions as mentioned in section~\ref{section:zoneprobas}, an important quantity is the summed probability $\cF$ that the clearance zone is in the first empty interval irrespective of the number of particles that passed. This quantity is shown in the inset of Fig\,\ref{fig:E}.

From the study of the $\cF_N(t)$ some properties of a system with infinite closing rate could be deduced. For instance, the probability that the clearance zone has closed at time $t$ would be equal to $\int_{s=0}^t f(s) \dd s$. 

\subsection{Scaling properties for large $N$}
\label{subsection:scalingN}

\begin{figure}
\begin{center}
\scalebox{0.5}
{\includegraphics{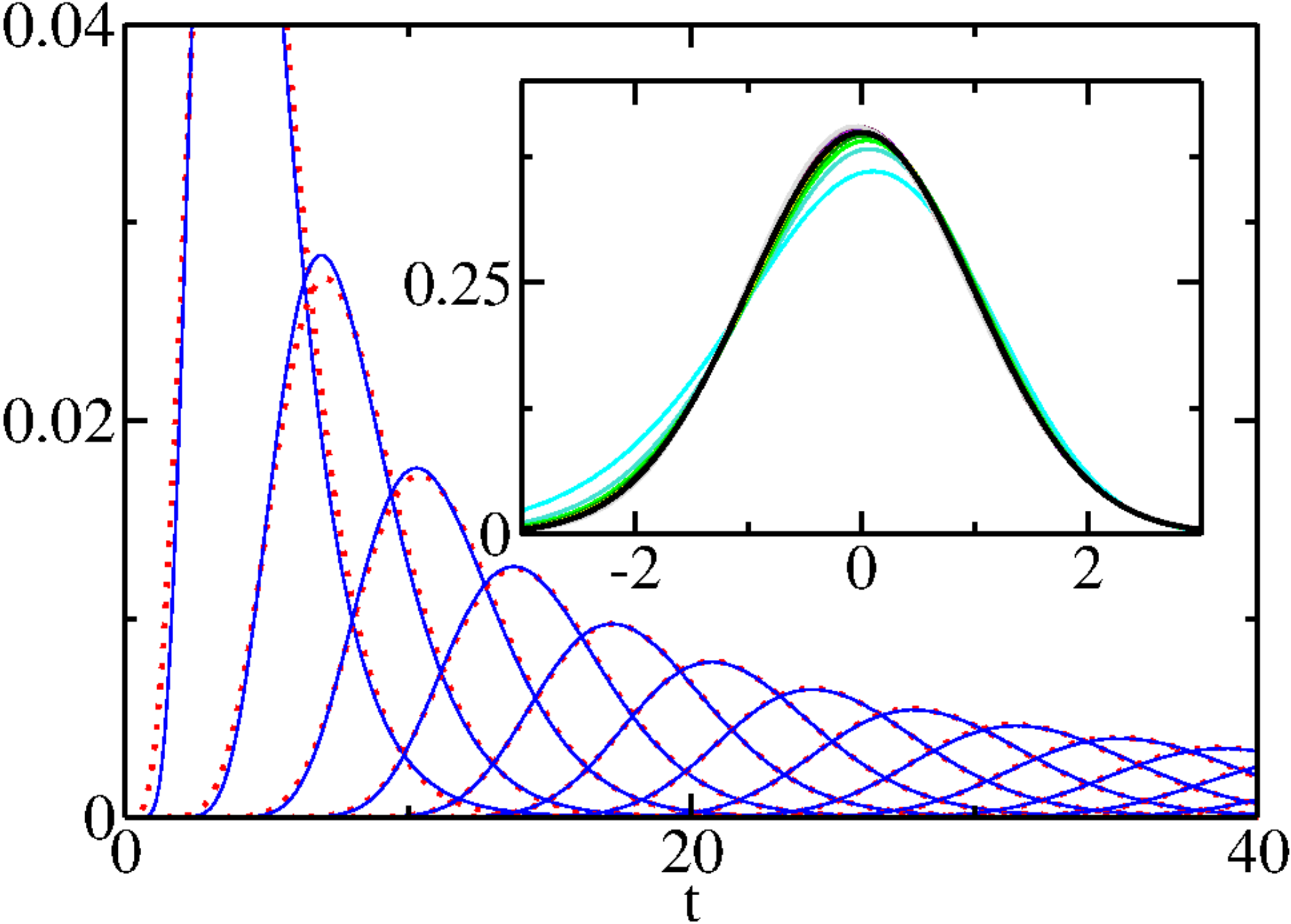}} 
\end{center}
\caption{\small Main curve: The functions $\cF_N(t)$ for $\nz=4$ and $N=1,\ldots,13$. The blue solid lines are the numerical curves and the red dashed lines are the log-normal fits~(\ref{eq:lognormal}). 
Inset: The  $13$ first $\cF_N$ after rescaling of the vertical axis by a factor $\frac{\sqrt{V_N} \ee^{M_N-V_N/2}}{\cN_N}$, as a function of $T$ (defined in the main text), compared to the standard Gaussian $\frac{1}{\sqrt{2 \pi}} \ee^{-\frac{x^2}{2}}$ (black solid line) corresponding to~(\ref{eq:lognormal}). The measured (colored) curves go up in the center as $N$ increases.}  
\label{fig:scalingFNnz4}
\end{figure}

In this subsection we study numerically the scaling behavior of the $\cF_N(t)$ for large $N$. The $\cF_N(t)$ are plotted for $\nz=4$ and $N=1,\ldots,13$ in Fig.\,\ref{fig:scalingFNnz4}.

\begin{figure}
\begin{center}
     \includegraphics[width=0.75\textwidth]{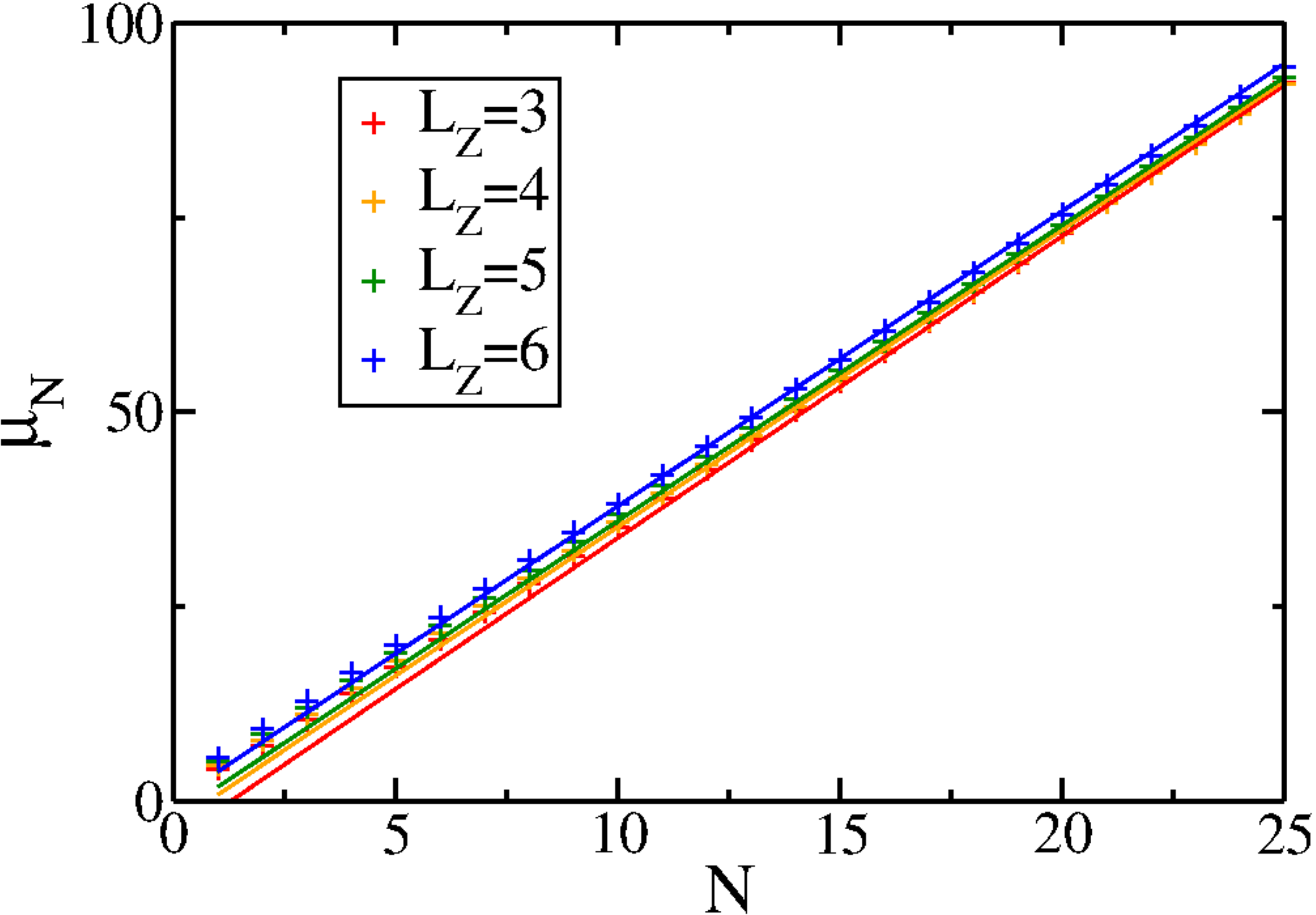} 
     \includegraphics[width=0.75\textwidth]{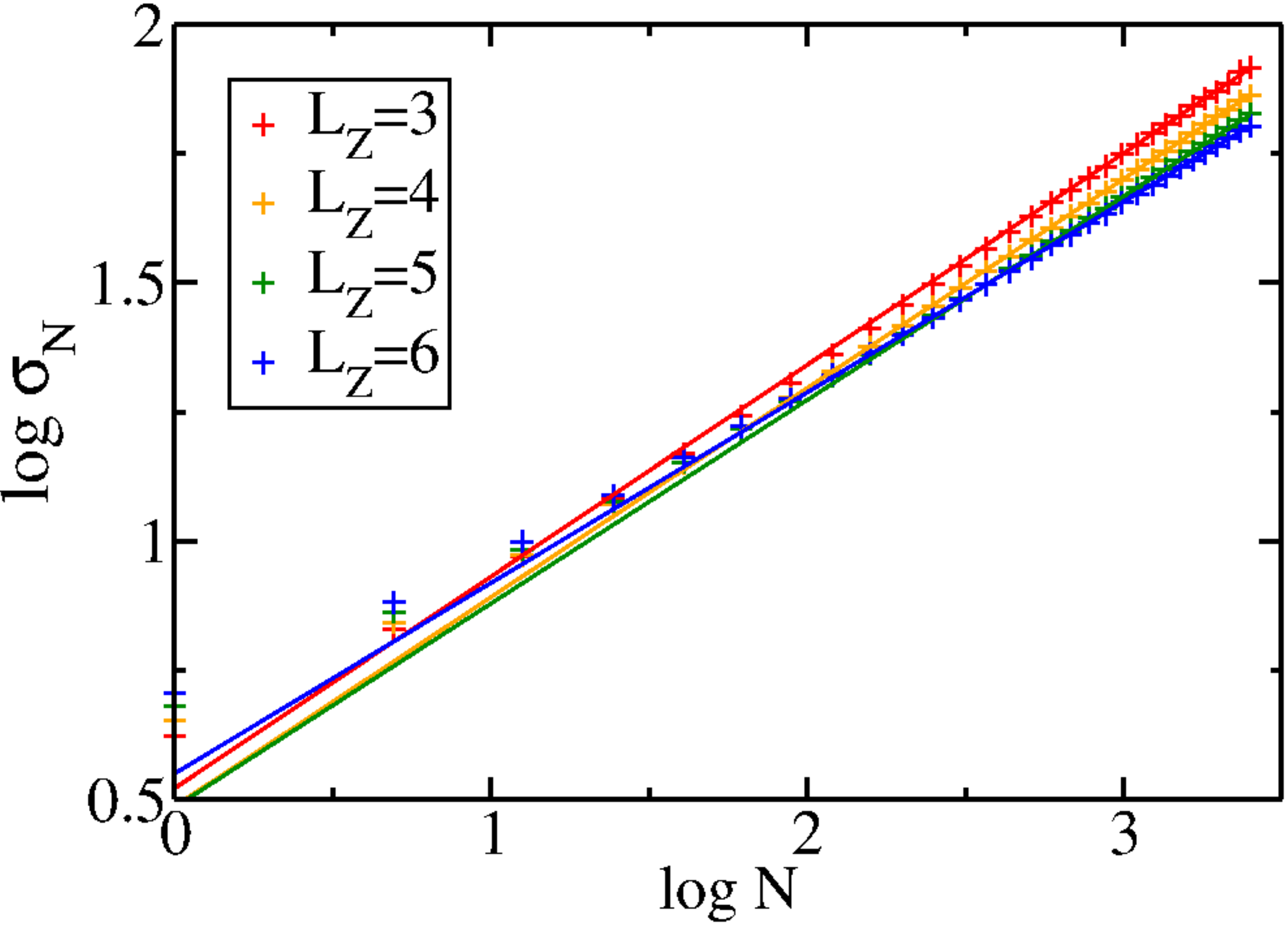} 
     \end{center}
\caption{\small Plot of  $\mu_N$ as a function of $N$ (top) and $\log \sigma_N$ as a function of $\log N$ (bottom) for $\nz$ between $3$ and $6$. The $+$ symbols are the numerical measurements and the solid lines are fits as described in the main text. Only the last $10$ points of every curve are fitted.
\label{fig:normmeanstd}
}
\end{figure}

We start by measuring the norm and the first cumulants of the $\cF_N(t)$ for a given value of $\nz$. We define
\begin{eqnarray}
 \label{eq:defNmusigma}
\cN_N &\equiv& \int_{t=0}^\infty \cF_N(t) \dd t, \nonumber \\
\mu_N &\equiv& \frac{1}{\cN_N} \int_{t=0}^\infty t \cF_N(t) \dd t, \\
\sigma_N &\equiv& \frac{1}{\cN_N} \int_{t=0}^\infty t^2 \cF_N(t) \dd t - \mu_N^2. \nonumber \\
\end{eqnarray}
We now study the dependency of the three observables in $N$, still for a given value of $\nz$. For large $N$, we see on Fig.\,\ref{fig:normmeanstd} that the mean increases linearly $\mu_N = b_1 N +b_2$ and the variance increases algebraically $\log \sigma_N = c_1 \log N + c_2$. 
The case of the norm is more complicated and the discussion is delayed. In the fits for $\mu_N$ and $\sigma_N$, the coefficients are of course expected to depend on $\nz$. Table \ref{table:scaling} shows the best fitting parameters for several $\nz$ values. We shall now give some elements of interpretation of these coefficients.

\begin{table}{}
\begin{center}
\begin{tabular}{c|cccc}
\label{table:scaling}
 $\nz$ & $b_1$ & $b_2$ & $c_1$ & $c_2$ \\
\hline 
3 & 3.88 & -5 & 0.410 & 0.52 \\
4 & 3.82 & -3 & 0.405 & 0.49 \\
5 & 3.80 & -2 & 0.394 & 0.48 \\
6 & 3.79 & 0 & 0.369 & 0.55 \\
\vdots & \vdots & \vdots & \vdots & \vdots \\
10 & 3.82 & 5 & 0.358 & 0.61 \\
\vdots & \vdots & \vdots & \vdots & \vdots \\
15 & 3.88 & 10 & 0.323 & 0.68 \\
\end{tabular}
\end{center}
\caption{Values of the fitting parameters for the first two moments of the $\cF_N(t)$ for various $\nz$ values.} 
\label{fig:tablepoles} 
\end{table}

\vspace{0.5cm}

For large $N$, the mean $\mu_N$ of $\cF_N$ can be estimated as follows.
The first particle typically needs $\nz+1$ time units to reach the end of $\Z$,
and after that we consider crudely that the current between sites $\nz$ and
$\nz+1$ is constant equal to $\jj$ between times $\nz+1$ and $\mu_N$, where we neglect the duration of the empty interval after the $N^\mathrm{th}$ particle exits. We therefore have $\jj \sim \frac{N-1}{\mu_N -
\nz}$, or $\mu_N \sim \frac{N}{\jj}+\nz-\frac{1}{\jj}$, giving $b_1 \sim
\jj^{-1}$ and $b_2 \sim \nz-\jj^{-1}$. This explains the order of magnitude of
$b_1$, and of $b_2$ when $\nz$ becomes large. Indeed, if we suppose that $\Z$
is in maximal current phase then $\jj=1/4$ and $b_1 \sim 4$, $b_2 \sim \nz-4$. The
numerical measurements show that $b_1 \lesssim 4$, and since $b_1$ can still be
interpreted as the inverse of the average current we have $\jj \gtrsim 1/4$, as
can be expected for the transient regime. The constant $b_2$ is a subdominant
term and is evaluated less precisely. We however notice that its variation with
$\nz$ is compatible with the preceding discussion at large $\nz$.

The values of $c_1$ and $c_2$ seem trickier to explain.
Numerically we have $c_1 < 1/2$, so that the standard
deviation is lower than for a sum of i.i.d. variables. There is a negative
feedback, that may come from the fact that when a particle does not hop for a
long time, thus reducing the current, a denser region appears behind it which
will create a current higher than expected afterwards.

The norms $\cN_N$ are defined as the time integrals of the $\cF_N(t)$. In the
approximation~(\ref{eq:defpin})-(\ref{eq:exprFf}),
	we have
\begin{eqnarray}
  \label{eq:demo1}
\cN_N & = & \int_{t=0}^\infty
  \int_{s=0}^t f_N(s) \ee^{-(t-s)} \dd s \dd t\nonumber \\
  & = & \int_{s=0}^\infty f_N(s)
  \int_{t=s}^\infty
  \ee^{-(t-s)} \dd t \dd s \nonumber \\
  & = & \int_{t=0}^\infty f_N(t) \dd t
\end{eqnarray}
Thus the $\cN_N$ may be directly interpreted
as the probabilities that the first empty interval occurs after the passage of
$N$ particles. More generally, $\cN_N$ is the product between the probability
that the first empty interval occurs after the passage of $N$ particles and the
mean duration $\langle T_N \rangle$ of this interval, as
shown in~\ref{section:interpnorm}

\begin{figure}
\begin{center}
\scalebox{0.3}
{\includegraphics{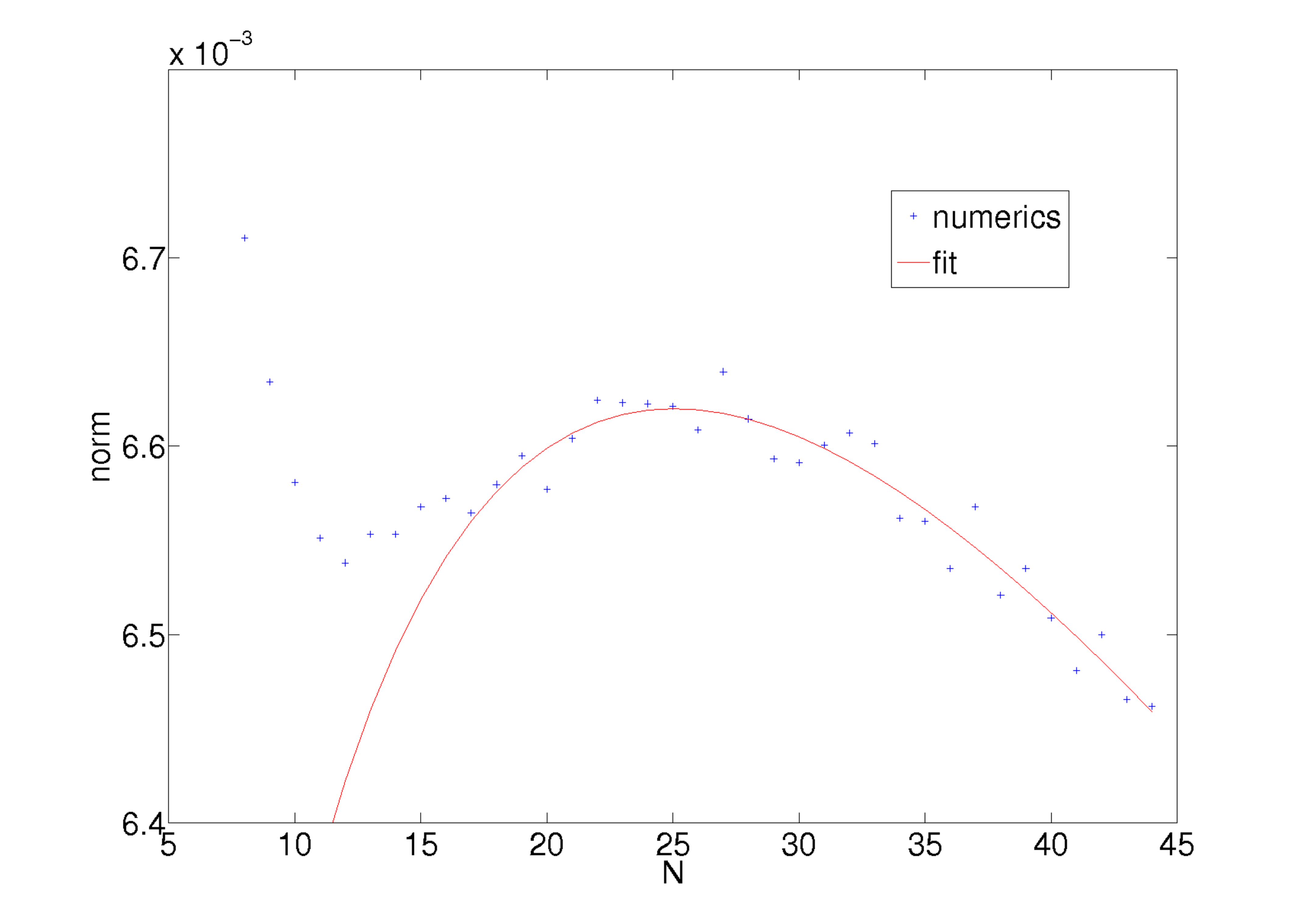}} 
\end{center}
\caption{\small Norm of $\cF_N(t)$ as a function of $N$ for $\nz=8$ obtained by Monte Carlo simulation (blue). The red curve is a fit of the form~(\ref{eq:exprnorm}), carried out over the points $N=25, \ldots, 44$.}  
\label{fig:normnz}
\end{figure}

The norms $\cN_N$ are rather complicated functions of $N$.
Naively one would expect that $\cN_N$ is a decreasing function of $N$, as
the zone $\Z$ is less and less likely not to have been empty before the $N^\mathrm{th}$ particle passes.
Actually it is a non-monotonous function, as seen on Fig.\,\ref{fig:normnz}. We would like to guess a form for the large $N$ behavior.
We have seen in subsection~\ref{subsection:sumN} that the probability that $Z$ becomes empty at time $t$ is
slightly less than $2^{-\nz}$, due to long-lived algebraic correlations.
In other words, the clearance zone is  slightly more likely to be empty for increasing large times, \textit{i.e.} for increasing large $N$.
Hence we take as a guess
\begin{eqnarray}
 \label{eq:exprnorm}
\cN_N &\sim& \Pr[\Z\mathrm{\; has\; never\; been\; empty\; at\; }t] \nonumber \\ &&\times \Pr[\Z\mathrm{\; becomes\; empty\; at\; }t] \big|_{t = \jj^{-1} N} \nonumber \\
&\sim& K_1\ee^{-K_2 t} \times \bigg( 2^{-\nz} - \frac{K_3}{t^{K_4}} \bigg) \big|_{t = \jj^{-1} N} \\
&=& K_1\ee^{-K_2 \jj^{-1} N} \bigg( 2^{-\nz} - \frac{K_3 \jj^{K_4}}{N^{K_4}} \bigg), \nonumber
\end{eqnarray}
where the $K_i$ are unknown constants. It can be
shown that the expression~(\ref{eq:exprnorm}) has a maximum.
Indeed this is confirmed by the numerical results obtained for $\nz \geq 8$
in Fig.\,\ref{fig:normnz}. Numerical results show that the correction term $\frac{K_3}{t^{K_4}}$ increases with the size of the bottleneck, which makes the maximum more and more prominent as $\nz$ increases.

Numerically, we find that the
shape of each $\cF_N(t)$ is very close to a log-normal distribution,
\begin{eqnarray}
  \label{eq:lognormal}
\cL_N(t) &\equiv& \frac{\cN_N}{t \sqrt{2 \pi V_N}} \ee^{- \frac{(\log t - M_N)^2}{2 V_N}}\nonumber\\
&=& \frac{\ee^{V_N/2 -M_N} \cN_N}{\sqrt{V_N}} \frac{1}{\sqrt{2 \pi}}\exp \bigg[ -\frac{1}{2} \bigg( \frac{\log t - M_N + V_N}{\sqrt{V_N}}\bigg)^2 \bigg],
\end{eqnarray}
where we ensure that the mean and the variance of $\cF_N$ and $\cL_N$ are the same by setting
\begin{eqnarray}
 \label{eq:MNVN}
M_N &=& \log \mu_N -\frac{1}{2} \log \bigg( 1 +\frac{\sigma_N^2}{\mu_N^2}\bigg),\nonumber\\
V_N &=& \log \bigg( 1 +\frac{\sigma_N^2}{\mu_N^2}\bigg).
\end{eqnarray}
To check how close to a log-normal distribution $\cF_N(t)$ is,
we rescaled the axes and plotted $\frac{\sqrt{V_N} \ee^{M_N-V_N/2}}{\cN_N} \cF_N(t)$ as a function of $T \equiv \frac{\log t}{\sqrt{V_N}} - \frac{M_N}{\sqrt{V_N}} + \sqrt{V_N}$ for every $N$. The result is shown on Fig.\,\ref{fig:scalingFNnz4} for $\nz = 4$. The curves for very low value of $N$ seem to get closer and closer to the Gaussian limit corresponding to~(\ref{eq:lognormal}),
but as $N$ grows the sequence of curves slightly overtakes the Gaussian and seems to converge towards a different limit. The log normal distribution however remains a very good approximation.

\section{Conclusion}
\label{section:ccl}

In this work we have taken a simple well-known system, the infinite TASEP with a step initial condition, and we have studied the probability that a certain clearance zone becomes empty. The quantities we have focused on are non-stationary, finite-time and finite-space quantities. An exact analytic expression was obtained for the probability $\cE_N(t)$ that the clearance zone is empty at a given time conditioned on the number of particles that passed the clearance zone. The sum of these contributions, \textit{i.e.} the probability that the clearance zone $\Z$ is empty, was however harder to obtain in a simple form. This latter quantity was shown to converge only algebraically to its equilibrium value, as long-lived spatial correlations survive for a very long time.

A step was made towards computing even more complicated, history-dependent quantities such as the $\cF_N(t)$, defined as the probabilities that the clearance zone is in its first empty interval at time $t$, with $N$ particles having exited it. While the $N=1$ case is special in that $\cF_1(t)$ does not truly depend on the history, we presented an exact expression for $\cF_2(t)$ in terms of the propagator. Using this expression we derived the asymptotic behavior of $\cF_2(t)$ for large times and showed that the exponential part was proportional to $\ee^{-t}$, a feature that we argued to be true for any $\cF_N(t)$. We have proposed an approximate recurrence relation linking the $\cF_N(t)$ to the already known $\cE_N(t)$, that works quite well for large enough values of $\nz$. Finally, we measured the main characteristics of the $\cF_N(t)$ and discussed their physical meaning.

This work can be seen as a first step in the study of more complex systems involving a clearance zone switching between open and closed states, in which closing can occur only if the clearance zone is empty~\cite{turci2013, jelic_a_s2012}. In this paper the clearance zone $\Z$ does not affect the dynamics of the particles. It would seem natural to enable the closing of $\Z$ with some rate, as in the model of Ref.\,\cite{turci2013}. The closing of the clearance zone in some realizations of the system history would however modify the probabilities $\cE_N$ in a way quite difficult to compute exactly as it would once again involve the past history of each realization. In the case of an infinite closing rate, our results already provide the distribution of closing times. 

\section*{Acknowledgements}

We thank T. Sasamoto and G. Schehr for useful discussions.

\appendix

\section{Expression of $\cF_1(t)$}
\label{section:exprF1}

The expression of $\cF^*_1(t)$ can be obtained by a reasoning similar to the one of subsection~\ref{subsection:combNe1}. The main difference is that particle $1$ must hop more than $\nz-1$ times between times $s$ and $t$.
\begin{eqnarray}
\label{eq:cFst1comb}
 \cE^*_1(t) &=& \cF^*_1(t)\nonumber\\ &=&\int_{s=0}^t \dd s \ee^{-s} (1-\ee^{-s}) \ee^{-2(t-s)} \sum_{k=\nz-1}^\infty \frac{(t-s)^k}{k!}\nonumber \\&&+ \int_{s=0}^t \dd s \ee^{-2s} \ee^{-2(t-s)} (1+t-s) \sum_{k=\nz-1}^\infty \frac{(t-s)^k}{k!} \\
&=& (2t-\nz) \ee^{-t} - (t-\nz) \frac{\Gamma(\nz,t)}{\Gamma(\nz)} \ee^{-t} \nonumber\\ &&-\frac{t \Gamma(\nz-1,t)}{\Gamma(\nz-1)} \ee^{-t} + \frac{t^{\nz}}{\Gamma(\nz+1)} \ee^{-2 t}, \nonumber
\end{eqnarray}
By combining~(\ref{eq:phiphist}) and~(\ref{eq:cFst1comb}) we also obtain
\begin{eqnarray}
\label{eq:cF1comb}
 \cF_1(t) &=& \int_{s=0}^t \ee^{-(t-s)} \cF^*_1(s) \dd s  \\
&=& \ee^{-t} t (t-\nz) - \frac{\ee^{-t}}{\Gamma(\nz)} \bigg[ \frac{\Gamma(\nz+2)-\Gamma(\nz+2,t)}{2} + \frac{t^2}{2} \Gamma(\nz,t) \nonumber\\ &&- \nz (\Gamma(\nz+1) - \Gamma(\nz+1,t))  - \nz t \Gamma(\nz,t)\bigg] \nonumber\\ &&- \frac{\ee^{-t}}{2 \Gamma(\nz-1)} [ \Gamma(\nz+1)-\Gamma(\nz+1,t) + t^2 \Gamma(\nz-1,t)] \nonumber\\ &&+ \frac{\ee^{-t}}{\Gamma(\nz+1)} (\Gamma(\nz+1)-\Gamma(\nz+1,t)) \nonumber.
\end{eqnarray}

\section{Large time asymptotics of $F_2(t)$}
\label{section:asymptF2}

We evaluate the two parts of~(\ref{eq:cF2det}) for large values of the time arguments $s$ and $t-s$. It can indeed be shown by examination that no dominant term comes from the parts where $s$ or $t-s$ are of order $1$. We first write
\begin{eqnarray}
 \label{eq:dvptG3}
G(\bx|\by;s) &=& F_0 (x_3-y_3;s)  F_0 (x_2-y_2;s)  F_0 (x_1-y_1;s) \nonumber\\ &&+ F_1 (x_2-y_3;s)  F_1 (x_1-y_2;s)  F_{-2} (x_3-y_1;s) \nonumber\\&&+ F_2 (x_2-y_3;s) F_{-1} (x_3-y_2;s)  F_{-1} (x_2-y_1;s)   \nonumber\\ && - F_0 (x_3-y_3;s)  F_{1} (x_1-y_2;s)  F_{-1} (x_2-y_1;s) \\&&-F_1 (x_2-y_3;s)  F_{-1} (x_3-y_2;s)  F_0 (x_1-y_1;s)  \nonumber\\ && -F_2 (x_1-y_3;s)  F_0 (x_2-y_2;s)  F_{-2} (x_3-y_1;s).\nonumber
\end{eqnarray}
Now remember (subsection \ref{subsection:prop}) that all the $F_n(x;s)$ with $n \leq 0$ are either zero or bring a factor $\ee^{-s}$ with them at long times, so that all the terms in the RHS decay like $\ee^{-2 s}$ except for the second one. We have the asymptotic equivalents $F_1(x;s) \sim 1$ and $F_{-2}(x;t) \sim \frac{s^{2+x}}{(2+x)!} \ee^{-s}$ for $x+2 \geq 0$. For large $s$ the propagator appearing in~(\ref{eq:cF2det}) then behaves asymptotically as
\begin{equation}
 \label{eq:G3asympt}
G_3((\nz,x_2^s,x_3^s)|\bx^0;s) = \frac{s^{2+x^s_3}}{(2+x^s_3)!} \ee^{-s} + o(\ee^{-s}),
\end{equation}
where $x^s_3 + 2 \geq 0$ for any time, as $x_3^0 = -2$. We use an analogous method to estimate the second part of the RHS of~(\ref{eq:cF2det}). For this second half the summations over $x_2^t$ and $x_3^t$ can be performed right away using again the identity~(\ref{eq:sumFn}). In principle the summation over $x_1^t$ could be performed as well, but we choose not to do so for the moment. We get
\begin{eqnarray}
 \label{eq:G3ppoasympt}
 &&\sum_{x_1^t = \nz+2}^{\infty} \sum_{x_2^t = \nz+1}^{x_1^f -1} G_3(\bx^t|(\nz+1,x_2^s,x_3^s);t-s) \\ &=& [C_1 (t-s)^{-x_3^s} \Theta(x_3^t \geq x_3^s) + \frac{(t-s)^{2-x_2^s}}{(x_3^t-x_2^s+1)!} \Theta(x_3^t \geq x_2^s-1) \nonumber\\ &&+ C_2 (t-s)^{3-\nz} \Theta(x_3^t \geq \nz-1)] \times (t-s)^{x_3^t} \ee^{-(t-s)} + o(\ee^{-(t-s)}),\nonumber
\end{eqnarray}
where $C_1$ and $C_2$ are two constants. The $\Theta$ factors come from the fact that the $F_n(x;t)$ become zero for $n \leq 0$ and $x \leq n-1$. The first condition $x_3^t \geq x_3^s$ is obviously always true, and the third one $x_3^t \geq \nz-1$ is always false for all the interesting cases $\nz \geq 2$. Combining the second condition with the summation intervals in Eq.\,(\ref{eq:cF2det}), only the term with $x^t_3=0$ and $x^s_2 = 1$ survives. After performing the remaining sums the second term is shown to dominate. In particular $x_3^s = 0$ is the dominant term because of the factor (\ref{eq:G3asympt}). This finally gives the result~(\ref{eq:cFdetasymp}).

\section{Interpretation of $\cN_N$}
\label{section:interpnorm}

Here we give a precise argument that proves that $\cN_N$ may be interpreted as the product of the probability that the first empty interval occurs after the passage of $N$ particles with the average duration of this interval, denoted $\langle T_N \rangle$.

We introduce the moment $t^E$ when the clearance zone $\Z$ is empty for the first time and the state of the system $\bx^E$ at this moment. We also use the shorthand $P^E(t,\bx)$ for the probability that $\Z$ is empty for the first time at time $t$ in the configuration $\bx$, and we write $\Pr[.|.]$ for the conditional expectation. Then $\cF_N(t)$ may be decomposed as follows,
\begin{eqnarray}
 \label{eq:decompFN}
\cF_N(t) &=& \dd t^E \int_{t^E=0}^t \sum^N_{\bx^E} 
 P^E(t^E, \bx^E) \nonumber \\ &&\times \Pr[\mathrm{No\;entrance\;in\;} \Z  \mathrm{\;in\;} (t^E;t) \;|\; (t^E,\bx^E) ],
\end{eqnarray}
where the superscript on the sum over $\bx^E$ indicates that the sum is restricted to configurations $\bx^E$ in which exactly $N$ particles have exited the clearance zone, $\bx^E_N  =\nz+1$ and $\bx^E_{N+1} < 0$.

Taking the integral over $t$, we can exchange the integrals
\begin{eqnarray}
\label{eq:exchangeint}
\int_{t=0}^\infty \cF_N(t) \dd t &=& \int_{t^E=0}^\infty \dd t^E \int_{t=t^E}^\infty \dd t \sum^N_{\bx^E} 
P^E(t^E, \bx^E) \nonumber \\ &&\times \Pr[\mathrm{No\;entrance\;in\;} \Z \mathrm{\;in\;} (t^E;t) \;|\; (t^E,\bx^E) ] \\
&=& \int_{t^E=0}^\infty \dd t^E \sum^N_{\bx^E} 
P^E(t^E, \bx^E) \nonumber \\ &&\times \int_{t=t^E}^\infty \dd t \Pr[\mathrm{No\;entrance\;in\;} \Z  \mathrm{\;in\;} (t^E;t) \;|\; (t^E,\bx^E) ]. \nonumber
\end{eqnarray}

We now use the fact that $\int_{t^E=0}^\infty \sum^N_{\bx^E} P^E(t^E,\bx^E)$ is equal to the probability that $\Z$ is empty for the first time after$N$ passages, which is also $\int_{t=0}^\infty f_N(t) \dd t$. Defining $u = t-t^E$, we get
\begin{eqnarray}
 \label{eq:decompNN}
 \cN_N &=& \int_{t=0}^\infty \cF_N(t) \dd t \\
        &=& \sum_{\bx^E}^N \left[ \frac{ P^E(\bx^E)}{\sum^N_{\by}  P^E(\by)} \int_{u=0}^\infty \Pr[\mathrm{No\;entrance\;in}\; \Z \mathrm{\;in\;} (0;u) |\; (0,\bx^E)]  \dd u \right] \nonumber \\ &&\times \int_{t=0}^\infty f_N(t) \dd t, \nonumber
\end{eqnarray}
where $P^E(\bx^E) = \int_{t^E = 0}^\infty P^E(t^E,\bx^E) \dd t^E$ is the probability that $\Z$ is empty for the first time in configuration $\bx^E$.
The factor $  \frac{P^E(\bx^E)}{\sum^N_{\by} P^E(\by) \dd s}$
	is the probability that $\Z$ is empty for the first time in the
	configuration $\bx^E$ conditioned by the fact that exactly $N$

Thus
the whole sum over $\bx^E$ is
equal to $\langle T_N \rangle$ and we may finally write 
\begin{equation}
 \label{eq:interpNN}
 \cN_N =  \langle T_N \rangle \int_{t=0}^\infty f_N(t) \dd t 
\end{equation}

In the recurrence of subsection~\ref{subsection:rec} we made the simplifying assumption $\Pr[\mathrm{First\;entrance\;in} \;\Z\; \mathrm{at}\; u\; |\; (0,\bx)]= \pin(u) = \ee^{-u}$, which gives $\langle T_N \rangle = 1$ for all $N$. This yields $\cN_N = 
\int_{t=0}^\infty f_N(t) \dd t 
=  \Pr[\Z\; \mathrm{empty\;for\;the\;first\;time\;after} \;N\; \mathrm{passages}]$

\vspace{0.5cm}

\end{document}